\documentclass[a4paper, submission, Phys]{SciPost}

\usepackage[utf8]{inputenc}
\usepackage{amsmath,amssymb,amsfonts} 
\usepackage{float}
\usepackage{braket}
\usepackage{tikz}
\usepackage{pgfplots}
\usepackage{hyperref}
\usepackage[caption=false]{subfig}
\usepackage{lineno}
\newcommand{\norm}[1]{\left| #1 \right|}
\newcommand{\Tr}{\mathrm{Tr}}
\newcommand{\dd}{\mathrm{d}}
\newcommand{\ee}{\mathrm{e}}
\newcommand{\expval}[1]{\langle #1 \rangle}
\newcommand{\eq}[1]{\begin{equation}#1\end{equation}}
\newcommand{\fop}{c} 
\newcommand{\mop}{m} 
\newcommand{\mtop}{\tilde{m}} 
\newcommand\chlength{L} 
\newcommand{\lopw}[1]{\mathcal{O}(w_{#1},\bar w_{#1})}
\newcommand{\lopdw}[1]{\mathcal{O}^\dagger(w_{#1},\bar w_{#1})}

\begin{document}

\begin{center}{\Large \textbf{
Entanglement spreading after local fermionic excitations\\ in the XXZ chain
}}\end{center}

\begin{center}
Matthias Gruber,
Viktor Eisler
\end{center}

\begin{center}
  Institute of Theoretical and Computational Physics,
  Graz University of Technology, NAWI Graz, Petersgasse 16, 8010 Graz, Austria
\\
\end{center}

\begin{center}
\today
\end{center}


\section*{Abstract}
{\bf
We study the spreading of entanglement produced by the time evolution of
a local fermionic excitation created above the ground state of the XXZ chain.
The resulting entropy profiles are investigated via density-matrix
renormalization group calculations, and compared to a quasiparticle ansatz.
In particular, we assume that the entanglement is dominantly
carried by spinon excitations traveling at different velocities, and the entropy
profile is reproduced by a probabilistic expression involving the
density fraction of the spinons reaching the subsystem. The ansatz
works well in the gapless phase for moderate values of the XXZ anisotropy,
eventually deteriorating as other types of quasiparticle excitations gain spectral weight.
Furthermore, if the initial state is excited by a local Majorana fermion, we observe a
nontrivial rescaling of the entropy profiles. This effect is further investigated
in a conformal field theory framework, carrying out calculations for the
Luttinger liquid theory. Finally, we also consider excitations creating an
antiferromagnetic domain wall in the gapped phase of the chain, and find
again a modified quasiparticle ansatz with a multiplicative factor.
}

\vspace{10pt}
\noindent\rule{\textwidth}{1pt}
\tableofcontents\thispagestyle{fancy}
\noindent\rule{\textwidth}{1pt}
\vspace{10pt}

\section{Introduction}

The non-equilibrium dynamics of integrable models has developed into a vast field of research
\cite{calabrese2016introduction}. Among the numerous aspects, the understanding of
local relaxation and equilibration in closed quantum systems has become a central topic of
investigation \cite{polkovnikov2011noneqclosed,gogolin2016equinclosed}. In this respect, integrable systems show a rather
peculiar behaviour, as the dynamics is characterized by the existence of stable quasiparticle
excitations. This is intimately related to the extensive number of nontrivial conservation laws,
which nevertheless allow for a local relaxation in a generalized sense \cite{VR16}.

Starting from the early studies of this topic, it was identified that the spreading of
entanglement must play a key role in our understanding of integrable dynamics.
Ground states of homogeneous, local Hamiltonians have a low amount of entanglement,
typically satisfying an area law \cite{eisert2010arealaw}. However, considering  the time evolution
with respect to a different Hamiltonian as in the context of a global quantum quench
\cite{essler2016quench}, the rapid \emph{linear} growth of entanglement was attributed to
the ballistic propagation of entangled quasiparticle pairs \cite{CC05}.
These quasiparticles transmit entanglement over large distances, contributing to the
buildup of an extensive entropy within any given subsystem, which signals
the onset of some local thermalization. Specifically, in one-dimensional integrable chains
it has been verified that the entanglement entropy accumulated in a subsystem actually
plays the role of the thermal entropy as described by the generalized Gibbs ensemble
\cite{AC17,AC18,calabrese2020entanglement}.

The global quench is the simplest representative of an initial state
that has an extensive amount of energy above the ground state of the
Hamiltonian governing the dynamics, thus acting as a reservoir of
quasiparticle excitations. The interpretation, however, becomes more complicated
if the initial state lies in the low-energy regime. A particular example is the
local quench, where the final Hamiltonian is disturbed only locally with respect
to the initial one, such as joining two initially separated quantum chains.
At criticality, the entanglement spreading can be captured via conformal field theory (CFT)
\cite{CC07,CC16,SD11}, predicting a slow \emph{logarithmic}
growth of the entropy, which was indeed observed in free-fermion chains
\cite{EP07}. However, despite signatures of the underlying quasiparticle
dynamics, such as a light-cone spreading with the maximal group velocity,
it is unclear how the individual quasiparticles contribute to the entropy.

Yet another situation that has been studied intensively within CFT is the
so-called local operator excitation \cite{nozaki2014qentanglement,he2014qudim,nozaki2014notes}.
Here the low-energy initial state is excited from the vacuum of the CFT by the insertion
of a local primary operator, while the Hamiltonian is left untouched. The disturbance
has then a linear propagation, increasing the entanglement of a segment only while passing
through it, with a \emph{constant} excess entropy determined by the quantum
dimension of the local primary \cite{nozaki2014qentanglement,he2014qudim,nozaki2014notes}.
The calculations have been extended in various directions, considering fermionic
\cite{nozaki2016quantent} or descendant fields \cite{caputa2015conffam,chen2015desclocalop},
multiple excitations \cite{guo2018multipleLO},
as well as the effects of finite temperatures \cite{caputa2015quantent} or boundaries \cite{guo2015renyi}.

Despite this increased attention, there have been much less studies on entanglement spreading 
after local excitations in integrable quantum chains. The CFT predictions have been tested
on the critical transverse Ising \cite{caputa2017qdim} and XX chains \cite{zhang2020subdistance},
for various local operators that are lattice analogs of primary or descendant fields. On the other hand,
entanglement spreading has also been considered in the non-critical ordered phase of the Ising
\cite{eisler2016universal} and XY chains \cite{eisler2018hydrophasetransXY,eisler2020frontdyn},
starting from a domain-wall initial state excited by a local Majorana operator.
Remarkably, the emerging profile of the excess entropy was shown to be captured by a simple
probabilistic quasiparticle ansatz \cite{eisler2020frontdyn}. Indeed, taking into account the
dispersive spreading of quasiparticles, only a certain fraction of the initially localized excitation
will cross the subsystem boundary located at a certain distance. Interpreting this quasiparticle
fraction as the probability of finding the excitation within the subsystem, the excess
entropy is simply given by a binary expression \cite{eisler2020frontdyn}.

Here we aim to extend the quasiparticle description of entanglement spreading to
local fermionic excitations in the XXZ chain. Being a Bethe ansatz integrable interacting
model \cite{Takahashi,Franchini}, its quasiparticle content is much more complex than in
the free-fermion systems considered so far. Nevertheless, since our local excitations
probe the low-energy physics, it seems reasonable that the dominant weight is carried
by low-lying spinon excitations, which we shall assume to build our quasiparticle
ansatz. Compared against the profiles of the excess entropy, as obtained from density-matrix
renormalization group (DMRG) calculations
\cite{white1992DMRG,white1993DMRG,schollwoeck2011density},
we observe a good agreement after a local fermion creation
for moderate values of the interaction. For larger interactions in the gapless phase,
one finds deviations that can be attributed to different types of quasiparticles with higher energy.

We also study the profiles after a local Majorana excitation, which seem to be given
by a simple rescaling of the spinon ansatz. This result is supplemented
by CFT calculations carried out for the Luttinger liquid theory, which describes
the low-energy physics of the XXZ chain. We find that, due to the left-right mixing
of the chiral bosonic modes, the asymptotic excess entropy is doubled for the
Majorana excitation, although with a very slow convergence towards this value.
Finally, in the gapped phase of the chain we study the excess entropy profile
after a local Majorana operator that excites an antiferromagnetic domain wall.
Here our numerical results suggest that the spinon ansatz is multiplied by a
nontrivial factor, related to the ground-state entropy.

The rest of the manuscript is structured as follows.
In section \ref{sec: XXZ_ex} we introduce the XXZ chain and discuss its low-lying excitations.
Section \ref{sec: gapless} is devoted to the study of entanglement spreading after local excitations
in the gapless phase: we first introduce a quasiparticle ansatz for the excess entanglement,
followed by our numerical studies of a fermion creation as well as a Majorana excitation.
Our results for the gapless regime are complemented by a calculation of the R\'enyi entropy
within a CFT framework in section \ref{sec: CFT}. Finally, in section \ref{sec: gapped}
we consider entanglement and magnetization profiles after a domain-wall excitation in the
gapped regime. Our closing remarks are given in section \ref {sec: summary}, followed by an appendix
containing the details of the CFT calculations.

\section{XXZ chain and low-energy excitations \label{sec: XXZ_ex}}
We consider an XXZ chain of length $\chlength$ with open boundary conditions that is given by the Hamiltonian
\begin{equation}
  H = J \sum_{j=-\chlength/2+1}^{\chlength/2-1} \left( S^x_j S^x_{j+1} + S^y_j S^y_{j+1} +
  \Delta S^z_j S^z_{j+1} \right) \, ,
  \label{eq: H}
\end{equation}
where $S^\alpha_j=\sigma^\alpha_j/2$ are spin-$1/2$ operators acting on site $j$, and $\Delta$ is the anisotropy.
The energy scale is set by the coupling $J$ which we fix at $J=1$.
The XXZ Hamiltonian \eqref{eq: H} conserves the total magnetization $S^z$ in $z$-direction
and we will be interested in its ground state in the zero-magnetization sector $S^z=0$.
Equivalently, the XXZ spin chain can be rewritten in terms of spinless fermions
by performing a Jordan-Wigner transformation, which brings  \eqref{eq: H} into the form
\begin{equation}
    H = \sum_{j=-\chlength/2+1}^{\chlength/2-1} \left[ \frac{1}{2}(\fop^\dagger_j \fop_{j+1} + \fop^\dagger_{j+1} \fop_j)
    + \Delta \left( \fop^\dagger_j \fop_j - \frac{1}{2} \right) \left( \fop^\dagger_{j+1} \fop_{j+1} - \frac{1}{2} \right) \right]\, ,
      \label{eq: Hf}
\end{equation}
where $\fop^\dagger_j$ ($\fop_j$) are fermionic creation (annihilation) operators,
satisfying anticommutation relations $\{\fop_i,\fop^\dagger_j\} = \delta_{ij}$.
One then has a half-filled fermionic hopping chain with nearest-neighbour interactions of strength $\Delta$.
For $\norm{\Delta} \le 1$ the system is in a critical phase with gapless excitations above the
ground state, whereas a gap opens for $\norm{\Delta} > 1$.
The case $\Delta = 1$ corresponds to the isotropic Heisenberg antiferromagnet.

In the following we give a short and non-technical introduction to the construction
of the ground state and low-lying excited states of the XXZ chain.
To keep the discussion simple, we shall rather consider a periodic chain, 
and focus on the behaviour in the thermodynamic limit $L \to \infty$.
The exact eigenstates of the XXZ chain can be found from Bethe ansatz \cite{Takahashi,Franchini}.
These are constructed as a superposition of plane waves, the so-called magnons, labeled by
their rapidities $\lambda_i$ which provide a convenient parametrization of the quasimomenta.
The allowed values of the rapidities follow from the Bethe equations, with real solutions
corresponding to spin-wave like states.
Complex solutions organize themselves into strings and correspond to bound states.

For $|\Delta|<1$ the half-filled ground state is obtained by occupying all the
allowed vacancies of the $\chlength/2$ real rapidities, thus forming a tightly packed Fermi sea.
Low-energy excitations in the $S^z=1$ sector are called spinons and are created by removing
a rapidity. This creates two holes in the Fermi sea, with all the remaining rapidities moving slightly
with respect to their ground-state values, and the energy difference can be calculated from this back-flow effect.
In the thermodynamic limit, the result can be found analytically and written directly in terms
of the quasimomenta $q_1$ and $q_2$ of the two spinons as \cite{Takahashi}
\begin{equation}
    \Delta E = \varepsilon_s(q_1) + \varepsilon_s(q_2)\, ,
    \label{eq: DE}
\end{equation}
where the spinon dispersion relation in the gapless regime with $\Delta = \cos(\gamma)$ is given by
\begin{equation}
	\varepsilon_s( q) = \frac{\pi}{2} \frac{\sin(\gamma)}{\gamma} \sin(q) \, .
	\label{eq: E_s}
\end{equation}
Note that spinons are always excited in pairs, with the individual momenta confined to $0 \leq q_{1,2} \leq \pi $.
The total momentum is then given by $q_1+q_2$, and due to the additivity of \eqref{eq: DE}
one actually has a band of excitation energies. In particular, the lower edge of the two-spinon band is
obtained by setting $q_2 = 0$ or $q_2 = \pi$, and thus simply corresponds to shifting the dispersion
in \eqref{eq: E_s} for $q>\pi$. The group velocity of the spinons can be directly obtained
from the derivative of the dispersion
\begin{equation}
	v_s(q) = \frac{\dd \varepsilon_s(q)}{\dd q} = \frac{\pi}{2} \frac{\sin(\gamma)}{\gamma} \cos(q) \, .
	\label{eq: v_s}
\end{equation}

Further low-energy excitations with $S^z=1$ can be created by removing a single rapidity from the real axis
and placing it onto the $\mathrm{Im} \, \lambda = \pi$ axis. The energy of this particle-hole excitation
can be obtained, similarly to the spinon case, from the back-flow equations of the rapidities and yields
the dispersion \cite{Takahashi}
\begin{equation}
	\varepsilon_{ph}(q) = \pi\frac{\sin(\gamma)}{\gamma}
    \left| \sin\left(\frac{q}{2}\right) \right| \sqrt{ 1 + \cot^2\left(\frac{\pi}{2} \left(\frac{\pi}{\gamma} - 1\right)\right) \sin^2\left(\frac{q}{2}\right) } \, .
    \label{eq: E_ph}
\end{equation}
However, in contrast to spinons, particle-hole excitations are not composite objects and
their momentum range is thus $0\le q<2\pi$.
Note that these spin-wave like excitations are only physical for $-1 < \Delta < 0$,
i.e. in case of attractive interactions.
For low momenta $q \to 0$, the dispersion relation Eq.~\eqref{eq: E_ph} approaches the one for spinons in Eq.~\eqref{eq: E_s}.
The group velocities of particle-hole excitations are obtained by taking the derivative of $\varepsilon_{ph}(q)$.
Interestingly, it was found that the maximum particle-hole velocity can exceed the maximum spinon
velocity only if the anisotropy satisfies $\Delta < \Delta^* \approx -0.3$, which was demonstrated in a particular
quench protocol \cite{dePaula2017spinonbound}.

Finally, we consider the gapped phase where we focus exclusively on the antiferromagnetic regime $\Delta > 1$,
with the standard parametrization $\Delta=\cosh \phi$.
For even $\chlength$ the ground state has $S^z=0$ and is again given by $\chlength/2$ magnons with real rapidities.
However, the allowed number of vacancies is now $L/2+1$, which allows to construct a slightly shifted Fermi sea.
In the Ising limit $\Delta \rightarrow \infty$, this yields an exact twofold degenerate ground state, given by
the linear combinations of the two N\'eel states
\begin{equation}
    \ket{\psi_\pm} =
    \frac{\ket{ \uparrow \downarrow \uparrow \downarrow \dots} \pm
    \ket{\downarrow \uparrow \downarrow \uparrow \dots}}{\sqrt{2}}.
    \label{eq: gs_ising}
\end{equation}
For finite $\Delta$, the two states $\ket{\psi_\pm}$ constructed this way are only quasi-degenerate,
with an energy difference decaying exponentially in the system size $\chlength$.
Considering the thermodynamic limit one can write
\begin{equation}
    \ket{\psi_\pm} =
    \frac{\ket{\psi_\uparrow} \pm \ket{\psi_\downarrow}}{\sqrt{2}},
    \label{eq: gs_af}
\end{equation}
where $\ket{\psi_\uparrow}$ and $\ket{\psi_\downarrow}$ correspond to ground states with
spontaneously broken symmetry, displaying antiferromagnetic ordering. In fact,
the bulk expectation value of the staggered magnetization can be calculated analytically as 
\cite{baxter1973Fmodel,izergin1999spontmagXXZ}
\begin{equation}
    \bra{\psi_\uparrow}\sigma^z_j\ket{\psi_\uparrow}=
    -\bra{\psi_\downarrow}\sigma^z_j\ket{\psi_\downarrow}=
    (-1)^j\prod_{n=1}^{\infty} \tanh^2(n\phi) \, .
    \label{eq: sz_staggered}
\end{equation}

The low-lying excitations in the gapped phase 
are given again by spinons, by creating two holes in the Fermi sea.
The excitation energy is still given by Eq.~\eqref{eq: DE},
with the dispersion in the gapped phase obtained as \cite{Takahashi}
\begin{equation}
	\varepsilon_s(q) = \frac{\sinh(\phi)}{\pi} K(u) \sqrt{1-u^2 \cos^2(q)}\, ,
	\label{eq: E_s2}
\end{equation}
where the complete elliptic integral of the first kind reads
\begin{equation}
    K(u) = \int_0^{\pi/2} \frac{\dd p}{\sqrt{1-u^2\sin^2(p)}}
\end{equation}
and the elliptic modulus $u$ satisfies
\begin{equation}
	\phi = \pi \frac{K(\sqrt{1-u^2})}{K(u)}\, .
\end{equation}
The spinon velocity is obtained from the derivative of \eqref{eq: E_s2} and reads
\begin{equation}
    v_s(q) = \frac{\sinh(\phi)}{\pi} K(u) \frac{u^2 \sin(q)\cos(q)}{\sqrt{1-u^2 \cos^2(q)}}\, .
    \label{eq: vs_gapped}
\end{equation}

\section{Entanglement dynamics in the gapless phase \label{sec: gapless}}

The goal of this section is to study the entanglement dynamics after
a particular class of excitations. Namely, we first initialize the chain in its gapless ground state $\ket{\psi_0}$,
which is then excited by an operator that is strictly local in terms of the creation/annihilation operators
$c_j^\dag$  and $c_j$ appearing in the fermionic representation \eqref{eq: Hf} of the XXZ chain.
The system is then let evolve freely and we are interested in the emerging 
entanglement pattern in the time-evolved state $\ket{\psi(t)}$. For a bipartition into
a subsystem $A$ and the rest of the chain $B$, this is characterized
by the von Neumann entropy
\begin{equation}
    S(t) = -\Tr \left[ \rho_A(t) \ln \rho_A(t) \right],
\end{equation}
with the reduced density matrix $\rho_A(t) = \Tr_B \, \rho(t)$ and $\rho(t)=\ket{\psi(t)}\bra{\psi(t)}$.
In particular, we consider the bipartition $A = [-L/2+1,r]$ and $B = [r+1,L/2]$
and study the entropy profiles
\begin{equation}
    \Delta S = S(t) - S(0)
    \label{eq: DS}
\end{equation}
along the chain by varying $r$, where $r = 0$ corresponds to the half-chain.
Note that by subtracting the ground-state entropy $S(0)$, we aim to extract
information about the excess entanglement created by a local excitation.

In the following subsections we first introduce an intuitive picture for the description
of the entanglement spreading in terms of the low-lying quasiparticle excitations
introduced in Sec. \ref{sec: XXZ_ex}. We then proceed to the numerical study
of  the entanglement profiles after exciting the ground state with a fermionic
creation operator, and compare the results to our quasiparticle ansatz.
In the last part we consider an excitation created by a local Majorana fermion operator.

\subsection{Entanglement spreading in the quasiparticle picture\label{sec: QP}}

Let us consider an excitation above the ground state of the XXZ chain by acting
with a fermion creation operator $c_j^\dag$. To capture the dynamics,
one would have to first decompose the initial local excited state in the eigenbasis of the Hamiltonian.
As discussed in the previous section, these eigenstates are described by quasiparticles
parametrized by their rapidities or quasimomenta. The entanglement properties of various
eigenstates in the XXZ chain were studied before in \cite{alba2009excited,molter2014boundstates},
whereas a systematic CFT treatment of low-energy excitations was introduced in
\cite{alcaraz2011lowECFT,berganza2012lowECFT}.
In the framework of free quantum field theory, a surprisingly simple result on quasiparticle excitations
was recently found in \cite{castro-alvaredo2018entcontent,castro-alvaredo2018entcontent2}.
Namely, the excess entanglement measured from the ground state was found to be completely
independent of the quasiparticle momenta, depending only on the ratio $p$ of the subsystem and
full chain lengths. Moreover, for quasiparticles described by a single momentum, the excess
entropy is given by a binary formula $\Delta S=-p\ln p -(1-p)\ln(1-p)$, which allows
for a simple probabilistic interpretation. Indeed, the ratio $p$ is just the probability of finding the
quasiparticle within the subsystem.

Motivated by these results, we now put forward a simple ansatz for the spreading of
entanglement after the local excitation. Under time evolution, the quasiparticles involved
in the decomposition of the initial state spread out with their corresponding
group velocities. However, our main assumption is that their contribution to
entanglement is still independent of the momentum. Furthermore, we shall also
assume that the dominant part of the entanglement is carried by the lowest-lying
spinon modes, and that a spatially localized excitation translates to a homogeneous
distribution of the momenta in the initial state. Under these assumptions we expect
that the entanglement profile at time $t\gg1$ and distance $r\gg 1$ from the excitation,
in the space-time scaling limit $\zeta=r/t$ fixed, is determined exclusively via
\begin{equation}
	\mathcal{N} = \int_{0}^{\pi} \frac{\dd q}{\pi}\ \Theta( v_s(q) - \zeta )\, ,
	\label{eq: N}
\end{equation}
where $\Theta(x)$ is the Heaviside step function and $v_s(q)$ is the spinon velocity.
In fact, this is nothing else but the fraction of the spinon modes with sufficient velocity to arrive
at the subsystem. The simple probabilistic interpretation of the entanglement then leads to
the binary entropy formula for the profile
\begin{equation}
	\Delta S = -\mathcal{N} \ln (\mathcal{N}) - (1-\mathcal{N}) \ln (1-\mathcal{N})\, .
	\label{eq: delS_N}
\end{equation}
In particular, for the gapless case considered here, inserting the expression \eqref{eq: v_s}
of the spinon velocity into \eqref{eq: N}, the spinon fraction can immediately be found as
\begin{equation}
	\mathcal{N} = \frac{1}{\pi}\arccos \left(\frac{\zeta}{v}\right),
	\label{eq: N2}
\end{equation}
where $v=v_s(0)$ denotes the maximal spinon velocity.

In summary, our simplistic ansatz \eqref{eq: delS_N} provides an interpretation
of the excess entropy based on the dispersive dynamics of the quasiparticle modes,
where $\mathcal{N}$ is the fraction of the initially localized excitation that arrives at the subsystem.
In fact, the very same ansatz has recently been suggested for the description of entanglement
spreading after local fermionic excitations in the XY chain, finding an excellent agreement with numerics
\cite{eisler2020frontdyn}. Note, however, that the XY chain is equivalent to a free-fermion model
and thus all the single-particle modes can exactly be included in $\mathcal{N}$. In contrast,
for the interacting XXZ chain, restricting ourselves to the spinon modes should necessarily
introduce some limitations to the quasiparticle ansatz, as demonstrated in the following subsection.

\subsection{Local fermionic excitation}
We continue with the numerical study of the excitation produced by
the fermionic creation operator $\fop^\dagger_j$. The fermion operators are related to
the spin variables via the Jordan-Wigner transformation
\begin{equation}
	\fop^\dagger_j = \left( \prod_{l = -\chlength/2+1}^{j-1} \sigma_l^z \right) \sigma^+_j\, ,\qquad
	\fop_j = \left( \prod_{l = -\chlength/2+1}^{j-1} \sigma_l^z \right) \sigma^-_j\, ,
	\label{eq: fop}
\end{equation}
where $\sigma^\alpha_j$ are the Pauli matrices and
$\sigma^\pm_j = \left(\sigma^x_j \pm i \sigma^y_j\right)/2$.
For simplicity, we shall only consider the case where the excitation 
is created by $\fop^\dagger_1$ in the middle of the chain.
The time-evolved state after the excitation is then given by
\begin{equation}
	\ket{\psi(t)} = N^{-1/2} \ee^{-iHt} \fop^\dagger_1 \ket{\psi_0}\, ,
\end{equation}
where $\ket{\psi_0}$ is the ground state and the normalization is given by
\begin{equation}
	N = \bra{\psi_0} \fop_1 \fop^\dagger_1 \ket{\psi_0} = 1/2
\end{equation}
as the ground state is half filled. The time evolution is actually implemented
via time-dependent DMRG (tDMRG) \cite{white2004tDMRG,daley2004tMDRG} in the spin-representation of the XXZ chain, by first carrying out the ground-state
search and applying the string operator \eqref{eq: fop} onto the MPS representation of $\ket{\psi_0}$. The calculations were performed using the ITensor
C++ library \cite{itensor} and a truncated weight of $10^{-9}$.\\

%
\begin{figure}[H]
  \centering
  \hspace*{-0.2cm}
  \includegraphics[width=0.48\textwidth]{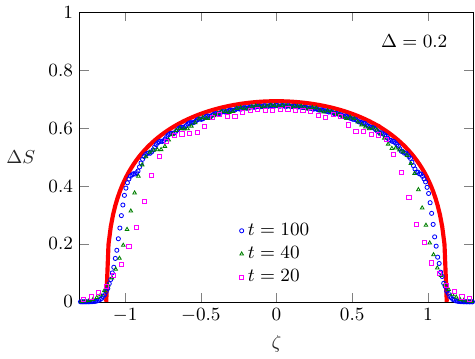}
  \hspace*{0.2cm}
  \includegraphics[width=0.48\textwidth]{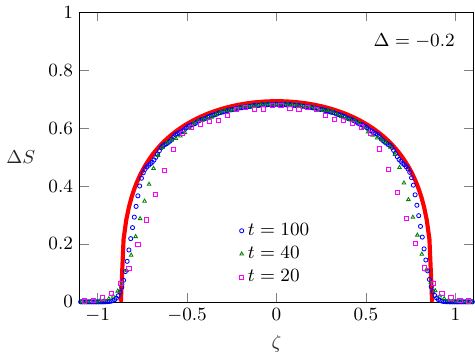}
  \hspace*{-0.2cm}
  \includegraphics[width=0.48\textwidth]{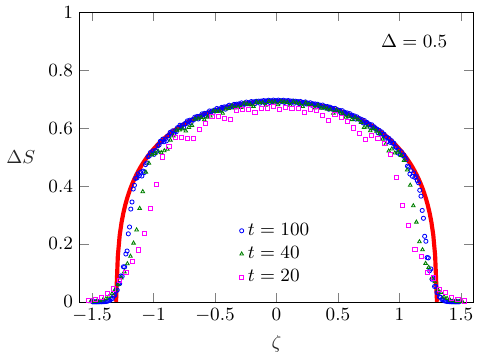}
  \hspace*{0.2cm}
  \includegraphics[width=0.48\textwidth]{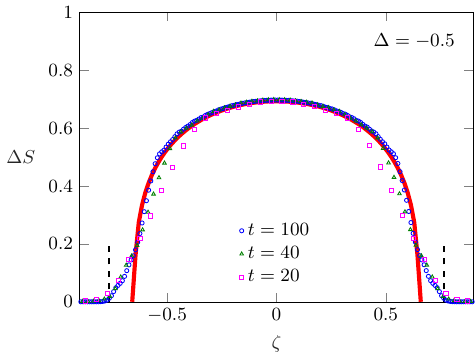}
  \hspace*{-0.2cm}
  \includegraphics[width=0.48\textwidth]{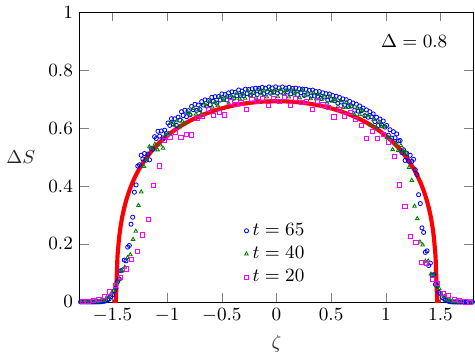}
  \hspace*{0.2cm}
  \includegraphics[width=0.48\textwidth]{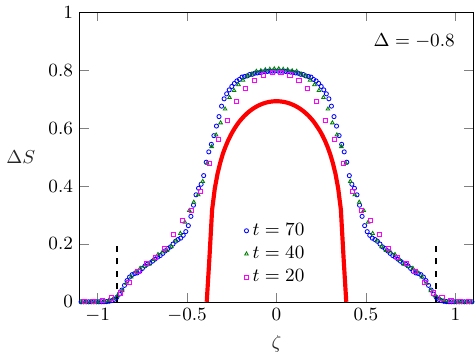}
  \caption{Excess entropy profiles $\Delta S$ obtained from tDMRG simulations at different times
  (symbols), after the excitation $\fop_1^\dagger$ in a chain of length $\chlength = 300$.
  The scaled profiles are plotted against $\zeta=r/t$ and compared to the quasiparticle ansatz (red lines)
  in Eq.~\eqref{eq: delS_N}.
  The dashed black lines denote the maximum velocity of the particle-hole excitations, derived from Eq.~\eqref{eq: E_ph}.}
  \label{fig: tDMRG_vs_QPk}
\end{figure}
%

The results of our simulations are shown in Fig. \ref{fig: tDMRG_vs_QPk} for various interaction strengths $\Delta$.
The different symbols correspond to snapshots of the entropy profile $\Delta S$ at different times,
plotted against the scaled distance $\zeta=r/t$. The quasiparticle ansatz \eqref{eq: delS_N}
computed using the spinon fraction \eqref{eq: N2} is shown by the red solid lines.
For moderate values of $|\Delta|$, one observes a very good agreement with the numerical profiles.
Systematic deviations from \eqref{eq: delS_N} occur for larger $\Delta$, especially in the attractive regime.
Indeed, for $\Delta=-0.5$ one already observes that the edges of the profile obtained from numerics
fall slightly outside of the spinon edge, whereas the bulk profile still shows a good agreement.
For $\Delta=-0.8$ the mismatch becomes more drastic both in the bulk and around the edges,
signaling the breakdown of the naive spinon ansatz. Clearly, for strong attractive interactions
the local excited state should have significant overlaps with other quasiparticle excitations of
the XXZ chain. In fact, as discussed in Sec. \ref{sec: XXZ_ex}, in this regime the maximum
velocity of particle-hole excitations exceeds the spinon velocity and matches perfectly the
edges of the profile, as indicated by the black dashed lines in Fig.~\ref{fig: tDMRG_vs_QPk}.
Hence, the entropy spreading should be determined by the coexistence of the
spinon and particle-hole excitations, allowing to reach values beyond $\ln(2)$.
Presumably, improving the ansatz \eqref{eq: delS_N} would require the knowledge of
the overlaps with the different families of quasiparticles. Finally, it should be noted that,
even though the edge locations of the profile seem to be captured, significant deviations in the bulk
also occur for large repulsive interactions (see $\Delta=0.8$ in Fig.~\ref{fig: tDMRG_vs_QPk}),
which might be due to bound-state contributions.

\subsection{Local Majorana excitation}
As a second example, we are going to consider local Majorana excitations,
given in terms of the spin variables via
\begin{equation}
    \mop_{2j-1} = \left( \prod_{l = -\chlength/2+1}^{j-1} \sigma^z_l \right) \sigma^x_{j}, \qquad
    \mop_{2j} = \left( \prod_{l = -\chlength/2+1}^{j-1} \sigma^z_l \right) \sigma^y_{j}\, ,
    \label{eq: majorana}
\end{equation}
and satisfying the anticommutation relations $\{\mop_k,\mop_l\} = 2 \delta_{kl}$.
Majorana operators are Hermitian and related to the fermion creation/annihilation operators as
$\mop_{2j-1} = \fop_j + \fop_j^\dagger$ and $\mop_{2j} = i\left( \fop_j - \fop_j^\dagger \right)$.
Focusing again on an excitation $\mop_1$ in the middle of the chain, 
the time-evolved stated is now given by
\begin{equation}
	\ket{\psi(t)} = \ee^{-iHt} \mop_1 \ket{\psi_0}\, .
	\label{eq: psit_mop}
\end{equation}

\begin{figure}[t]
  \centering
  \hspace*{-0.2cm}
  \includegraphics[width=0.48\textwidth]{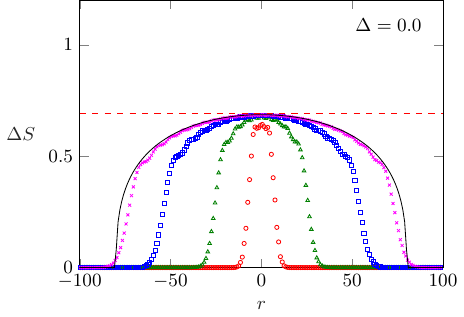}
  \hspace*{0.2cm}
  \includegraphics[width=0.48\textwidth]{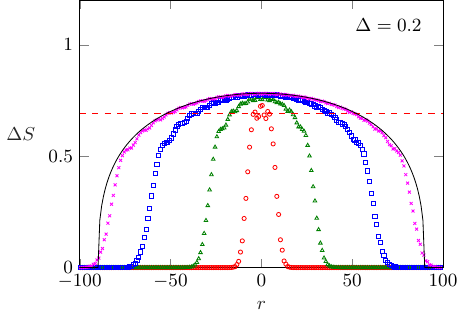}\\
  \vspace{0.4cm}
  \hspace*{-0.2cm}
  \includegraphics[width=0.48\textwidth]{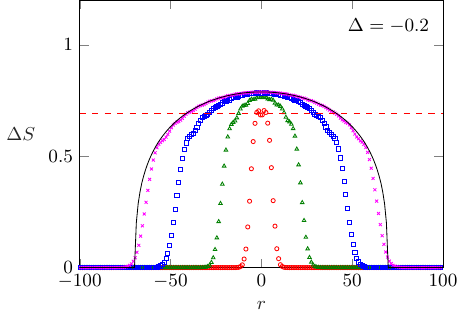}
  \hspace*{0.2cm}
  \includegraphics[width=0.48\textwidth]{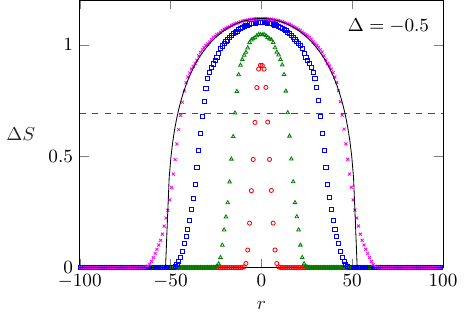}
  \caption{Excess entropy profiles $\Delta S$ as a function of $r$ at times $t = 10,30,60,80$ (red, green, blue, magenta)
  after the Majorana excitation $\mop_1$ for four different values of $\Delta$ and $\chlength = 200$.
  The red dashed lines indicate the value $\ln (2)$. The black solid lines show the spinon ansatz Eq.~\eqref{eq: delS_N}
  for $t=80$, multiplied by a constant to match the maxima of the profiles.}
  \label{fig: JW_S1_allbond}
\end{figure}

The entanglement profiles $\Delta S$ obtained from tDMRG simulations of \eqref{eq: psit_mop} are
depicted in Fig.~\ref{fig: JW_S1_allbond} for four different values of $\Delta$. To visualize the spreading
of the profile, we now plot the unscaled data against the location of the subsystem boundary.
For $\Delta=0$, the profile looks similar to that of the corresponding $c^\dag_1$ excitation
and is indeed perfectly reproduced by the quasiparticle ansatz \eqref{eq: delS_N}.
However, in the interacting case $\Delta \ne 0$, one observes a marked difference
when compared to the corresponding panels in Fig. \ref{fig: tDMRG_vs_QPk}.
Namely, the profiles in Fig. \ref{fig: JW_S1_allbond} clearly exceed the value $\ln(2)$,
indicated by the dashed horizontal lines, which is the maximum of the ansatz \eqref{eq: delS_N}.
Nevertheless, we observe that the profiles after the $m_1$ excitations can be well
described by a simple rescaling of the spinon ansatz \eqref{eq: delS_N}, as shown by the
solid lines in Fig. \ref{fig: JW_S1_allbond}. The constant factor multiplying the ansatz
is chosen such that the maxima of the profiles at $r=0$ are correctly reproduced.
%
%

\begin{figure}[ht]
  \centering
  \includegraphics[width=0.48\textwidth]{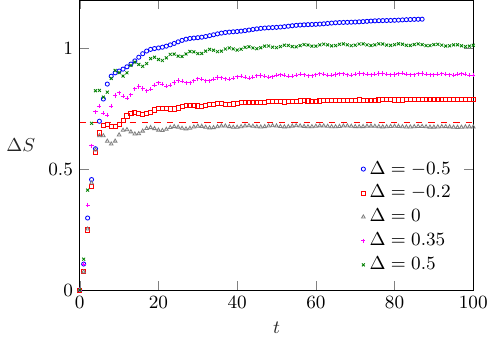}
  \includegraphics[width=0.48\textwidth]{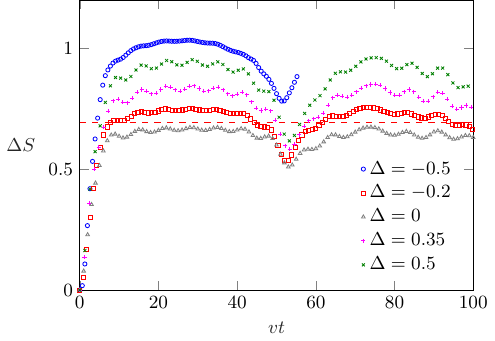}
  \caption{Left: Entropy growth in the middle of the chain $r = 0$,
  after the Majorana excitation $\mop_1$  for different values of $\Delta$ and $\chlength = 200$.
  The red dashed line indicates the value $\ln (2)$. Right: $\Delta S$ for a smaller chain with $L=50$
  against the scaled time $vt$ for the same $\Delta$ values.}
  \label{fig: JW_plateau}
\end{figure}

To better understand the behaviour of the maxima, on the left of Fig.~\ref{fig: JW_plateau} we plot
the time evolution of the excess entropy $\Delta S$ in the middle of the chain $(r=0)$ with $L=200$
and for various $\Delta$.
One observes that the asymptotic value of the excess entropy grows with increasing $|\Delta|$,
approaching its maximum very slowly in time. In fact, for even larger times the entropy
starts to decrease again as one approaches $vt \approx L$, when the fastest spinons
leave the subsystem after a reflection from the chain end. This is demonstrated on the
right of Fig.~\ref{fig: JW_plateau} by repeating the calculations for a smaller chain with $L=50$.
The emergence of a plateau is clearly visible, which then immediately repeats itself for
$vt>L$  due to the symmetry of the geometry, with the spinons reflected from the other
end of the chain entering the subsystem again. For an excitation located farther away
from the chain center, the corresponding plateaus would be separated.
However, the question why the height
of the plateau depends on the interaction strength $\Delta$ can only be answered via
a more involved CFT analysis of the problem, which is presented in the next section.

%
%
\section{Entanglement after local excitations in CFT \label{sec: CFT}}
The low-energy physics of the gapless XXZ chain can be captured within quantum field theory
via the bosonization procedure \cite{Senechal}. Using the fermionic
representation \eqref{eq: Hf} of the chain, one introduces the Heisenberg
operators $ c(x,\tau) = \ee^{\tau H} c_x \, \ee^{-\tau H}$,
where $x$ is the spatial coordinate along the chain and
we introduced the imaginary time $\tau=it$. Linearizing the dispersion around
the Fermi points, one can approximate
\begin{equation}
    c(x,\tau) \simeq \ee^{i k_F x} \psi(x,\tau) + \ee^{-i k_F x} \bar{\psi}(x,\tau)\, ,
\end{equation}
where $\psi (x,\tau)$ and $\bar\psi(x,\tau)$ are the right and left-moving components
of a fermion field. The phase factors with the Fermi momentum, where $k_F=\pi/2$
for a half-filled chain, are included to ensure that the chiral fermions are described by slowly varying fields.
Introducing the complex coordinates $w = v\tau - ix$ and $\bar w = v\tau + ix$, where
$v$ denotes the Fermi velocity, they can be written in a bosonized form \cite{Senechal}
\begin{equation}
    \psi(w) = \frac{1}{\sqrt{2\pi}} \ee^{-i \sqrt{4 \pi} \varphi(w)} \, , \qquad 
    \bar{\psi}(\bar{w}) = \frac{1}{\sqrt{2\pi}} \ee^{i \sqrt{4 \pi} \bar{\varphi}(\bar{w})}\, ,
    \label{eq: psibos}
\end{equation}
where $\varphi(w)$ and $\bar\varphi(\bar w)$ are the chiral boson fields.
In terms of the new bosonic variables
\begin{equation}
    \phi = \varphi + \bar{\varphi}\, , \qquad \theta = \varphi - \bar{\varphi}\, ,
    \label{eq: phitheta}
\end{equation}
one can show that the bosonized form of the XXZ chain \eqref{eq: Hf}
is described by the Luttinger liquid Hamiltonian \cite{Giamarchi}
\begin{equation}
    H_{LL} = \frac{v}{2} \int \dd x \left[ K(\partial_x \theta)^2 + K^{-1}(\partial_x \phi)^2 \right]\, .
    \label{eq: HLL}
\end{equation}
Apart from the velocity $v$, the Hamiltonian \eqref{eq: HLL} is characterized by
the Luttinger parameter $K$. Both of them can be fixed from the exact Bethe ansatz
solution as
\begin{equation}
    v = \frac{\pi}{2}\frac{\sin(\gamma)}{\gamma}\,, \qquad
    K = \frac{1}{2} \left(1 - \frac{\gamma}{\pi} \right)^{-1},
    \label{eq: K}
\end{equation}
with the usual parametrization $\Delta = \cos (\gamma)$. Note that $v=v_s(0)$ is just
the maximum of the spinon velocity \eqref{eq: v_s}.

In CFT language, the Luttinger liquid corresponds to a free compact boson field theory.
In order to study entanglement evolution after local operator excitations, we shall thus use 
the framework developed for a generic CFT \cite{nozaki2014qentanglement,he2014qudim}.
In the following we summarize the main steps of the procedure. Let us
consider the state
\begin{equation}
    \ket{\psi} = N^{-1/2} \mathcal{O}(-d)\ket{0}
\end{equation}
excited from the CFT vacuum $\ket{0}$ by insertion of the local operator $\mathcal{O}(-d)$,
where $N$ accounts for the normalization of the state. For the sake of generality,
we consider the situation where the excitation is inserted at a distance $d$ measured
from the center of the chain. After time evolution, the density matrix reads
\begin{equation}
    \rho(t) = N^{-1} \ee^{-iHt} \ee^{-\epsilon H} \mathcal{O}(-d) \ket{0}
    \bra{0} \mathcal{O}^\dagger(-d) \ee^{-\epsilon H} \ee^{iHt},
\end{equation}
where $\epsilon$ is a UV regularization that is required for the state to be normalizable. 
Working in a Heisenberg picture, the time evolution can be absorbed into the operators,
and the state can be represented as
\eq{
\rho(t) = \frac{\lopw{2} \ket{0} \bra{0} \lopdw{1}}{\braket{\lopdw{1} \lopw{2}}},
\label{eq: rhot}}
where the complex coordinates of the operator insertions are given by
\eq{
\begin{split}
&w_1 =  - i(vt - d) + \epsilon \, , \qquad \bar w_1 = - i(vt + d) + \epsilon \, , \\
&w_2 =  - i(vt - d) - \epsilon \, , \qquad \bar w_2 =  - i(vt + d) - \epsilon\,  .
\label{eq: w12}
\end{split}}
It should be stressed that the $\bar w_j$ coordinates are actually not the
complex conjugates of $w_j$, as we are assuming $\tau =it$ to be real,
such that we can work with Euclidean spacetime.

With the expression \eqref{eq: rhot} at hand, one can proceed to construct
the path-integral representation of the reduced density matrix, by opening
a cut at $\tau=0$ along the spatial coordinates of the subsystem $A$.
The R\'enyi entropy
\begin{equation}
    S_n(t) = \frac{1}{1-n} \ln \Tr \left[ \rho_A^n(t)\right]
\end{equation}
for integer $n$ can then be obtained by applying the replica trick \cite{CC09},
i.e. sewing together $n$ copies of the path integrals cyclically along the cuts.
In turn, one can express the excess R\'enyi entropy $\Delta S_n = S_n(t)-S_n(0)$
via correlation functions of the local operator as \cite{nozaki2014qentanglement,he2014qudim}
\begin{equation}
    \Delta S_n = \frac{1}{1-n} \log \left[ \frac{ \expval{\mathcal{O}^\dagger(w_1,\bar{w}_1)
    \mathcal{O}(w_2,\bar{w}_2)\dots \mathcal{O}(w_{2n} \bar{w}_{2n})}_{\Sigma_n}}
    {\expval{ \mathcal{O}^\dagger(w_1,\bar{w}_1) \mathcal{O}(w_2,\bar{w}_2) }^n_{\Sigma_1}} \right] ,
    \label{eq: DSn}
\end{equation}
where $\Sigma_n$ denotes the $n$-sheeted Riemann surface, with $w_{1},\dots,w_{2n}$
and $\bar w_{1},\dots,\bar w_{2n}$ being the replica coordinates of the insertion points \eqref{eq: w12}.

Although the expression \eqref{eq: DSn} for the excess R\'enyi entropy is very general,
the calculation of $2n$-point functions on the complicated Riemann surface $\Sigma_n$
may become rather involved. However, if the subsystem $A$ is given by a single interval
$0 \leq x \leq \ell$ in an infinite chain, the geometry can be simplified by the conformal transformation
\begin{equation}
    z = \left( \frac{w}{w+i\ell} \right)^{1/n}, \qquad
    \bar{z} = \left( \frac{\bar{w}}{\bar{w}-i \ell} \right)^{1/n},
\end{equation}
which maps the $n$-sheeted surface onto a single Riemann sheet.
This transformation leads to the holomorphic coordinates of the operator insertions
\begin{equation}
z_{2j-1} = e^{2 \pi i j/n} \left( \frac{d-vt-i\epsilon}{\ell+d-vt-i\epsilon} \right)^{1/n}, \qquad
z_{2j} = e^{2 \pi i j /n} \left( \frac{d - vt + i\epsilon}{\ell+d-vt+i\epsilon} \right)^{1/n},
\label{eq: zi}
\end{equation}
while the anti-holomorphic ones are given by
\begin{equation}
\bar{z}_{2j-1} = e^{-2 \pi i j/n} \left( \frac{d+vt+i\epsilon}{\ell+d+vt+i\epsilon} \right)^{1/n}, \qquad
\bar{z}_{2j} = e^{-2 \pi i j /n} \left( \frac{d + vt - i\epsilon}{\ell+d+vt-i\epsilon} \right)^{1/n}\, .
\label{eq: zib}
\end{equation}
Furthermore, if the local operators are primary fields of the CFT with respective
conformal dimensions $h_{\mathcal{O}}$ and $\bar{h}_\mathcal{O}$, the $2n$-point
function transforms as
%
\begin{align}
    \expval{ &\prod_{j=1}^n \mathcal{O}^\dagger(w_{2j-1},\bar{w}_{2j-1}) \mathcal{O}(w_{2j},\bar{w}_{2j}) }_{\Sigma_n} = \nonumber \\
    &\prod_{i=1}^{2n} \left( \frac{\dd w}{\dd z} \right)_{z_i}^{-h_{\mathcal{O}}} \left( \frac{\dd \bar{w}}{\dd \bar{z}} \right)_{\bar{z}_i}^{-\bar{h}_{\mathcal{O}}}
    \expval{ \prod_{j=1}^n \mathcal{O}^\dagger(z_{2j-1},\bar{z}_{2j-1}) \mathcal{O}(z_{2j},\bar{z}_{2j}) }_{\Sigma_1}\, .
    \label{eq: conf_trans}
\end{align}
%
In the end, one is left with a problem of calculating $2n$-point functions on the complex plane.
For the sake of simplicity, in the following we shall only consider the case $n=2$, and apply
the procedure outlined above to the Luttinger liquid theory, with the local excitations
considered in section \ref{sec: gapless}.

\subsection{Fermionic excitation}
We start with the fermion creation operator, which after bosonization \eqref{eq: psibos}
corresponds to the field insertion
\begin{equation}
    \mathcal{O}_f(w,\bar w) = \ee^{i k_F d} \ee^{i \sqrt{4 \pi} \varphi(w)} +
    \ee^{-i k_F d} \ee^{-i \sqrt{4 \pi} \bar\varphi(\bar w)} \, ,
\end{equation}
where we omitted normalization factors that cancel in the expression \eqref{eq: DSn}.
Clearly, $\mathcal{O}_f(w,\bar w)$ is not itself a primary operator but rather a linear combination of two. 
Hence, the calculation of the four-point function that appears in $\Delta S_2$ involves
a number of terms with primaries, each of which can be mapped from $\Sigma_2$
to the complex plane using the transformation rule \eqref{eq: conf_trans}.
The calculation of these correlation functions can be facilitated by first performing
a canonical transformation
\begin{equation}
     \qquad \theta' = \sqrt{K} \theta\, , \qquad \phi' = \frac{1}{\sqrt{K}} \phi\, .
\end{equation}
which absorbs the Luttinger parameter $K$ in the Hamiltonian \eqref{eq: HLL}.
However, since the variables $\theta$ and $\phi$ are actually linear combinations
\eqref{eq: phitheta} of the chiral bosons, the change of variables corresponds to
the Bogoliubov transformation
\begin{equation}
    \varphi = \cosh(\xi) \varphi' + \sinh(\xi) \bar{\varphi}' \qquad
    \bar{\varphi} = \sinh(\xi) \varphi' + \cosh(\xi) \bar{\varphi}'\, ,
    \label{eq: bt}
\end{equation}
where $K = \ee^{2 \xi}$. Thus, the transformation of the Luttinger liquid Hamiltonian
induces a left-right mixing of the chiral bosonic modes.
In the following we shall use the shorthand notations $c = \cosh(\xi)$ and $s = \sinh(\xi)$.

Clearly, our task now boils down to evaluate correlation functions of vertex operators
\begin{equation}
    V_{\alpha,\beta}(z,\bar{z}) = \ee^{i \sqrt{4\pi} \alpha \varphi'(z) + i\sqrt{4\pi} \beta \bar{\varphi}'(\bar{z})}
    \label{eq: vop}
\end{equation}
on the complex plane with respect to the Luttinger liquid theory scaled to the free-fermion point.
The $n$-point function of vertex operators is then well known and given by \cite{CFTbook}
\begin{equation}
    \expval{\prod_{i=1}^n V_{\alpha_i,\beta_i}(z_i,\bar{z}_i)} = \prod_{i<j} (z_i-z_j)^{\alpha_i \alpha_j}
    (\bar{z}_i - \bar{z}_j)^{\beta_i \beta_j},
    \label{eq: vertexcorr}
\end{equation}
where the neutrality conditions
\eq{
\sum_{i=1}^{n} \alpha_i=0 \, , \qquad
\sum_{i=1}^{n} \beta_i=0 \,  \qquad
\label{eq: neut}}
must be satisfied, otherwise the correlator vanishes. In particular, considering the two-point
function one immediately sees that the vertex operator \eqref{eq: vop} is a primary with scaling dimensions
$h=\alpha^2/2$ and $\bar h=\beta^2/2$.

With all the ingredients at hand, performing the calculation for $\Delta S_2$ is a straightforward
but cumbersome exercise, and we refer to Appendix~\ref{sec: CFT_appendix} for the main details.
It turns out that the result depends only on the cross-ratios
\begin{equation}
    \eta = \frac{z_{12} z_{34}}{z_{13} z_{24}}\, , \qquad
    \bar \eta = \frac{\bar z_{12} \bar z_{34}}{\bar z_{13} \bar z_{24}}
    \label{eq: crossr}
\end{equation}
of the holomorphic and anti-holomorphic coordinates \eqref{eq: zi} and \eqref{eq: zib},
where $z_{ij} = z_i - z_j$ and $\bar z_{ij} = \bar z_i -  \bar z_j$, respectively.
In terms of the cross-ratios, the final result reads
\begin{equation}
    \Delta S_2 = -\ln \left( \frac{1 + \left| \eta \right|^{(c+s)^2} + \left| 1-\eta \right|^{(c+s)^2}}{2} \right)\, .
    \label{eq: DS2f}
\end{equation}
It is important to stress that the notation $|\eta|$ should be understood as
$\left( \eta \bar{\eta} \right)^{1/2}$, since the two cross ratios are not conjugate variables.
In particular, in the limit $\epsilon \to 0$ of the regularization, one has the behaviour
\cite{nozaki2014qentanglement,he2014qudim}
\begin{equation}
    \lim_{\epsilon \to 0} \eta = 
    \begin{cases}
        0 &\text{if $0 < vt < d$ or $vt>d + \ell$}\\
        1 &\text{if $d < vt < d + \ell$}
    \end{cases}, \qquad
    \lim_{\epsilon \to 0} \bar{\eta} = 0\, .
    \label{eq: eta_lim}
\end{equation}
This yields the following limit for the R\'enyi entropy
\begin{equation}
    \lim_{\epsilon \to 0} \Delta S_2 = 
    \begin{cases}
        0 &\text{if $0 < vt < d$ and $vt>d + \ell$}\\
        \ln(2) &\text{if $d < vt < d + \ell$}
    \end{cases}\, .
    \label{eq: DS2flim}
\end{equation}

The result has a very simple interpretation. Namely, our excitation is an equal superposition
of a left- and right-moving fermion, and the entanglement is changed by $\ln(2)$ only when
the right-moving excitation is located within the interval. In fact, this is exactly the same
picture that lies behind the quasiparticle ansatz \eqref{eq: delS_N}, without the dispersion
of the wavefront. Interestingly, apart from the presence of the spinon velocity $v$,
the limiting result \eqref{eq: DS2flim} is independent of the anisotropy $\Delta$.
The only effect of the left-right boson mixing appears in the exponents of the
cross-ratios in \eqref{eq: DS2f}, which simply determines how the sharp step-function
for $\Delta S_2$ is rounded off for finite UV regularizations. In fact, this result
is very similar to the one obtained for a non-chiral EPR-primary excitation in Ref.
\cite{he2014qudim,caputa2015conffam}. Moreover, this is also a simple generalization
of the result in Ref. \cite{zhang2020subdistance}, where the superposition of purely
holomorphic and anti-holomorphic primaries was considered.

Finally, one should note that a numerical evaluation of $\Delta S_2$ in the XXZ
chain gives rise again to profiles that exceed $\ln(2)$ for larger values of $|\Delta |$,
similarly to what was observed for $\Delta S$ in Fig. \ref{fig: tDMRG_vs_QPk}.
The discrepancy is due to the fact that the Luttinger liquid theory describes only the
spinon modes and cannot account for other families of quasiparticles in the XXZ chain
that are responsible for the enhanced excess entropy.

\subsection{Majorana excitation}
We move on to consider the Majorana excitation
\begin{equation}
    \mathcal{O}_m(w,\bar{w}) =  \mathcal{O}_f(w,\bar{w}) + \mathcal{O}^\dagger_f(w,\bar{w}) \, .
\end{equation}
The calculation of $\Delta S_2$ follows the exact same procedure as for $\mathcal{O}_f(w,\bar{w})$,
however, one has now an even larger number of terms to consider. The main steps are
again outlined in Appendix~\ref{sec: CFT_appendix}, which lead to the result
\begin{equation}
	\Delta S_2 = -\ln \left( \frac{2A + B + C}{8} \right),
	\label{eq: DS2_CFT_m}
\end{equation}
where the terms in the logarithm are given by
\begin{align}
&A = \left| 1-\eta \right|^{(c+s)^2} + \left| 1-\eta \right|^{(c-s)^2} +  \left| \eta \right|^{(c+s)^2} + \left| \eta \right|^{(c-s)^2}\, 
	\label{eq: A}\\
&B = 2 + \eta^{2c^2} \bar{\eta} ^{2s^2}  + \eta^{2s^2} \bar{\eta} ^{2c^2}  + 
         \left(1-\eta \right)^{2c^2} \left(1-\bar{\eta} \right)^{2s^2} + \left(1-\eta \right)^{2s^2} \left(1-\bar{\eta} \right)^{2c^2}
         \label{eq: B} \\
&C = \left[ \left| \eta \right|^{(c+s)^2} \left| 1-\eta \right|^{(c-s)^2} +
         \left| \eta \right|^{(c-s)^2} \left| 1-\eta \right|^{(c+s)^2} \right](Z + \bar Z)
         \label{eq: C}
\end{align}
and a new variable is introduced as
\begin{equation}
	Z = \frac{z_1\bar z_2(1-\bar z_1^2)(1-z_2^2) }{\bar z_1 z_2 (1-z_1^2) (1-\bar z_2^2)} \, .
\end{equation}

The result is thus rather involved and cannot be written as a function of the cross-ratios alone.
However, in the limit $\epsilon \to 0$, the factors in $A$, $B$, and $C$ can trivially be evaluated
using \eqref{eq: eta_lim}, as well as using $Z \to 1$ and $\bar Z \to 1$. For the case $\Delta \ne 0$, this
leads to the following simple result
\begin{equation}
    \lim_{\epsilon \to 0} \Delta S_2 = 
    \begin{cases}
        0 &\text{if $0 < vt < d$ and $vt>d + \ell$}\\
        2\ln(2) &\text{if $d < vt < d + \ell$}
    \end{cases}\, .
    \label{eq: DS2mlim}
\end{equation}
In sharp contrast, for $\Delta=0$, where $c=1$ and $s=0$, one recovers the result \eqref{eq: DS2flim}.
Hence, one arrives at the rather surprising result that the excess entropy is doubled
in case of interactions, which must be a consequence of the left-right boson mixing.

\begin{figure}[H]
  \centering 
    \includegraphics[]{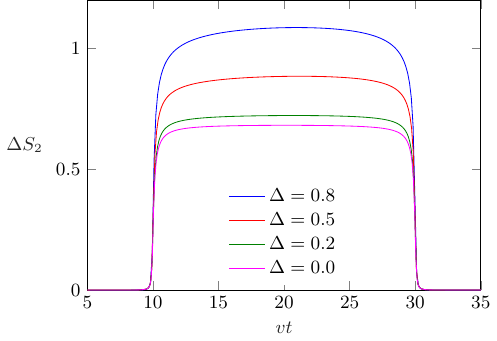}
  \caption{Time evolution of the excess R\'enyi entropy in Eq.~\eqref{eq: DS2_CFT_m}
  after the Majorana excitation with $\ell = 20$, $d = 10$ and $\epsilon = 0.1$.}
  \label{fig: CFT}
\end{figure}

Obviously, for finite values of the regularization $\epsilon$, this transition should take place
continuously, rather than giving an abrupt jump. The behaviour of $\Delta S_2$ for $\epsilon=0.1$
is shown in Fig.~\ref{fig: CFT} for an interval of length $\ell=20$ at a distance $d=10$
from the excitation. One can clearly see the development of a plateau for times $d < vt < d + \ell$,
the height of which increases monotonously with $\Delta$. Nevertheless, even for the largest
value $\Delta=0.8$, the expected maximum of $2\ln(2)$ is by far not reached.
The very slow convergence towards the $\epsilon \to 0$ (or, equivalently, $t \to \infty$) limit can
be understood by looking at the structure of the terms appearing in \eqref{eq: DS2_CFT_m}.
In fact, for smaller values of $|\Delta|$, the slowest converging pieces are given by
$\eta^{2c^2} \bar{\eta} ^{2s^2}$ as well as $\left(1-\eta \right)^{2s^2} \left(1-\bar{\eta} \right)^{2c^2}$
in the expression \eqref{eq: B} of $B$, due to the large-time behaviour
$\bar \eta \approx 1-\eta \approx (\epsilon/2vt)^2$ for $d \ll vt \ll \ell+d$.
Hence, the apparent nontrivial values of
the plateau in Fig. \ref{fig: CFT} is a consequence of the very slow decay $(\epsilon/vt)^{4s^2}$,
where the exponent for e.g. $\Delta=0.5$ is given by $4s^2 \approx 0.08$.
Clearly, observing convergence towards $\Delta S_2\to 2\ln(2)$ would require enormous
time scales as well as interval lengths.
%
%

Despite the different geometry considered for the CFT calculations,
we expect that the result \eqref{eq: DS2_CFT_m} should also give quantitative predictions
for the finite XXZ chain in a certain regime. First of all, for the half-chain
bipartition where the excitation is applied directly at the boundary, the role of the dispersion
should not play an important role, as all the excitations can immediately
enter the subsystem. Furthermore, one could argue that the finite chain
effectively corresponds to an interval of size $\ell=L$, which is the distance
the quasiparticles have to cover before leaving the subsystem after reflection
from the chain end. Clearly, the exact form of the plateau will not be the same
in the two cases, but one expects the CFT results to be applicable in a regime
$vt \ll L$. Finally, there is a highly nontrivial symmetry $s \to -s$ displayed by
all the terms \eqref{eq: A}-\eqref{eq: C} in the expression of $\Delta S_2$,
corresponding to a change of the Luttinger parameter $K \to 1/K$, which
is expected to be observed also in the lattice calculations.
Note that since $K=1$ corresponds to the free-fermion point $\Delta=0$,
the symmetry relates interaction strengths of different sign.
%
\begin{figure}[H]
  \centering
  \hspace*{-0.4cm}
  \includegraphics[height=110pt]{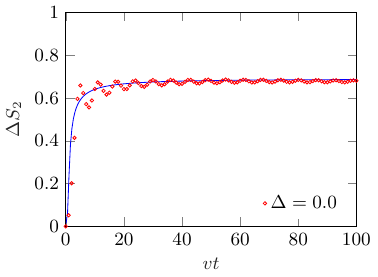}
  \hspace*{-0.4cm}
  \includegraphics[height=110pt]{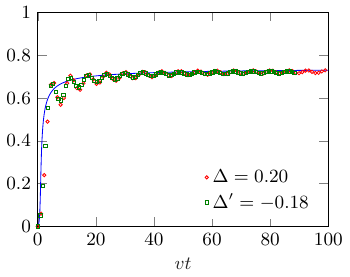}
  \hspace*{-0.4cm}
  \includegraphics[height=110pt]{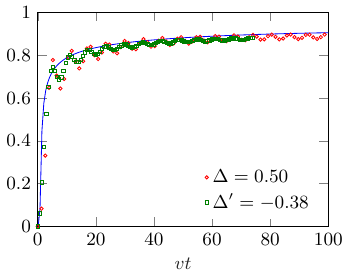}
  \caption{Growth of the R\'enyi entropy $\Delta S_2$ for pairs of conjugate interaction parameters
  $\Delta$ and $\Delta'$ (red and green symbols) for a chain of length $\chlength = 200$.
  The blue solid lines show the CFT result Eq.~\eqref{eq: DS2_CFT_m} with $\ell = 200$ and $d = 1$.
  The regularization $\epsilon = 0.55,0.40,0.35$ (from left to right) was tuned to obtain the best match with the tDMRG data.}
  \label{fig: CFT2}
\end{figure}
%
In Fig.~\ref{fig: CFT2} we show a comparison of $\Delta S_2$ obtained from
tDMRG calculations for a XXZ chain with $L=200$ divided in the middle, to the
CFT result \eqref{eq: DS2_CFT_m} shown by the blue solid lines. For the latter
we have set $\ell=L$ and $d=1$ as discussed above, whereas the regularization
$\epsilon$ was set by hand in order to achieve the best agreement with the
numerical data. One indeed observes that the CFT result gives, up to oscillations,
a good quantitative description of the XXZ numerics. Furthermore, for each
$\Delta \ne 0$, we also performed the calculation for the conjugate
$\Delta'$ corresponding to $K' = 1/K$, leading to a remarkably good collapse
of the curves.
\section{Entanglement dynamics in the gapped phase \label{sec: gapped}}
The CFT studies of the previous section give a rather good qualitative description
of the entanglement spreading in the critical phase of the XXZ chain.
To obtain a complete picture, in this section we shall study the dynamics in the
gapped antiferromagnetic phase. For a physically motivated setting,
we choose one of the symmetry-broken ground states $\ket{\psi_\uparrow}$
from Eq.~\eqref{eq: gs_af}, with a nonvanishing staggered magnetization \eqref{eq: sz_staggered}.
We now consider local Majorana operators, defined in terms of the spin variables as
\begin{equation}
    \mtop_{2j-1} = \left( \prod_{l = -\chlength/2+1}^{j-1} \sigma^x_l \right) \sigma^z_{j}\, , \qquad
    \mtop_{2j} = \left( \prod_{l = -\chlength/2+1}^{j-1} \sigma^x_l \right) \sigma^y_{j}\, .
    \label{eq: majorana2}
\end{equation}
Note that these operators differ from the ones in \eqref{eq: majorana} discussed in the gapless phase
by an interchange of the $x$ and $z$ spin components, but they also obey Majorana fermion statistics
with anticommutation relations $\{\mtop_k,\mtop_l\} = 2 \delta_{kl}$. We focus on the case of a
domain wall created by $\mtop_1$ in the center of the chain, which is then time evolved
by the XXZ Hamiltonian \eqref{eq: H}
\begin{equation}
	\ket{\psi(t)} = \ee^{-iHt} \mtop_1 \ket{\psi_\uparrow}\, .
	\label{eq: psit_gapped}
\end{equation}
Note that, in order to find the proper symmetry-broken ground state, 
in the DMRG simulation we add to the Hamiltonian a small staggered field in the $z$-direction,
which is then decreased towards zero during the sweeps.

\begin{figure}[H]
  \centering
  \hspace{-0.2cm}
  \includegraphics[height=150pt]{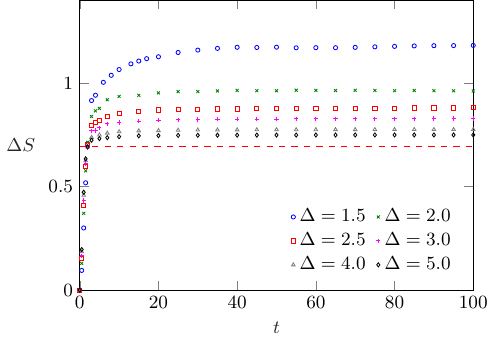}
  \hspace{0.2cm}
  \includegraphics[height=150pt]{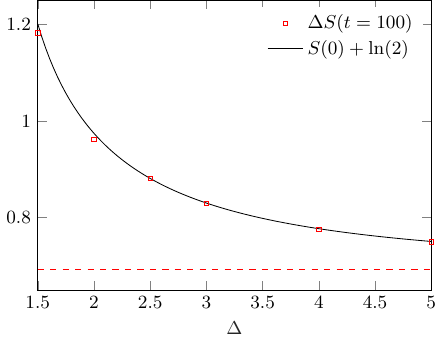}
  \caption{Left: Entanglement growth in the middle of the chain after a domain-wall excitation
  $\mtop_1$ for different values of $\Delta > 1$ and $\chlength = 400$.
  Right: $\Delta S$ at $t = 100$ compared to $S(0) + \ln(2)$ from Eq.~\eqref{eq: Sf1}. 
  The red dashed line denotes $\ln(2)$. Note the different vertical scales.}
  \label{fig: S_plateau}
\end{figure}

First we have a look at the entropy growth $\Delta S$ for the half-chain $r=0$ as a function of time,
shown on the left of Fig.~\ref{fig: S_plateau} for several values of the anisotropy $\Delta>1$.
One observes a clear saturation of the excess entropy for large times, which is reached very
quickly for large values of $\Delta$. The asymptotic value of $\Delta S$ decreases with $\Delta$
and always exceeds $\ln(2)$. Remarkably, as shown on the right of Fig. \ref{fig: S_plateau},
we find that the asymptotic excess entropy is well described by the formula $\Delta S = S(0) + \ln(2)$,
where $S(0)$ is the ground-state entropy of the half-chain in the symmetry-broken state.
Repeating the calculation for the excess R\'enyi entropy $\Delta S_2$, we find the exact
same relation with $S_2(0)$.

To gain a deeper understanding of the above relation, one should invoke the
exact results for the reduced density matrix of the half-chain, which can
can be found with the corner transfer matrix (CTM) method as \cite{peschel1999CTM}
\begin{equation}
    \rho_A = \frac{\ee^{-H_{CTM}}}{\Tr\left( \ee^{-H_{CTM}} \right)}\, , \qquad
     H_{CTM} = \sum_{j=0}^\infty \epsilon_j n_j\, ,
     \label{eq: CTM}
\end{equation}
where the single-particle eigenvalues are given by $\epsilon_j = 2 j \phi$ with
$\phi = \mathrm{acosh}(\Delta)$, and $n_j=0,1$ denotes fermionic occupation numbers.
In other words, the entanglement Hamiltonian $H_{CTM}$ of the ground state is characterized
by an equispaced single-particle entanglement spectrum. Strictly speaking, this result
applies to a half-infinite chain, but in practice it holds also for finite chains of length
much larger than the correlation length. Note also, that the result \eqref{eq: CTM}
applies for the symmetric ground state, whereas for the symmetry-broken state
the term $j=0$ is missing from the sum. In that case, the von Neumann and R\'enyi entropies
can be simply expressed as \cite{calabrese2010corrections}
\begin{equation}
    S(0) = \sum_{j=1}^{\infty} \left[ \log \left( 1 + \ee^{-2j\phi}\right) + 
    \frac{2 j\phi}{1 + \ee^{2 j \phi}} \right],
    \label{eq: Sf1}
\end{equation}
as well as
\begin{equation}
    S_n(0) = \frac{1}{1-n} \left[ \sum_{j=1}^{\infty} \log \left( 1 + \ee^{-2 n j \phi} \right) -
    n \sum_{j=1}^{\infty} \log \left( 1 + \ee^{-2 j \phi} \right) \right].
    \label{eq: S_n(0)}
\end{equation}
It is easy to see that the inclusion of the term $j=0$ with $\epsilon_0=0$ simply
yields an extra $\ln(2)$ contribution to the entropies. This change alone, however, would not explain
our findings for the asymptotic excess entropy in Fig.~\ref{fig: S_plateau}, which seems to
indicate that $S(t) \approx 2S(0)+\ln(2)$ for $t \gg 1$. Indeed, in order to obtain such a
formula, one would have to add a double degeneracy for each $\epsilon_j$ with $j \ne 0$.
Let us now discuss how such a degeneracy is reflected in the eigenvalues $\lambda_l$
of the reduced density matrix. In fact, it is more convenient to introduce the scaled quantity
\begin{equation}
    \nu_l = -\frac{1}{\phi}\ln \left( \frac{\lambda_l}{\lambda_0} \right),
    \label{eq: nul}
\end{equation}
where $\lambda_0$ denotes the maximal eigenvalue. For the initial symmetry-broken
ground state, $\nu_l$ are independent of $\Delta$ and can only assume even integer values,
with occasional multiplicities due to different integer partitions. The lowest lying $\lambda_l$
yield $\nu_l=0,2,4,6,6,\dots$, i.e. the first degeneracy appears as $6=2+4$.
The inclusion of the $\epsilon_0=0$ term simply gives an overall double degeneracy
of the levels $\lambda_l$. The doubling of the $\epsilon_j$ for $j \ne 0$
further increases the degeneracies. Altogether, the combined effect would lead
to the multiplicities $(2, 4, 6)$ for $\nu_l=0,2,4$.

\begin{figure}[H]
  \centering
  \includegraphics[]{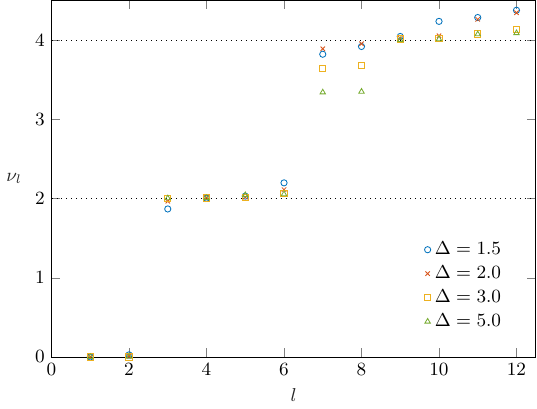}
  \caption{Scaled levels $\nu_l$ obtained from the reduced density matrix eigenvalues $\lambda_l$
   at time $t = 100$ via Eq.~\eqref{eq: nul} for different $\Delta$.}
  \label{fig: nul}
\end{figure}

To check these predictions, in Fig.~\ref{fig: nul} we have plotted the 12 lowest lying $\nu_l$
calculated from the reduced density matrix eigenvalues, as obtained from tDMRG simulations
after time evolving the state \eqref{eq: psit_gapped} to $t=100$. One can see that the
$\nu_l$ lie indeed rather close to the expected even integer values, approximately
reproducing the expected multiplicity structure. Interestingly, the largest deviation around
$\nu_l=4$ is found for $\Delta=5$, where one actually finds the best agreement with
the entropy formula, see Fig.~\ref{fig: S_plateau}. In fact, however, the contribution of these
eigenvalues to the entropy is already negligible. Note that the situation for larger values of
$\nu_l$ is much less clear, as they correspond to very small eigenvalues $\lambda_l$
which are already seriously affected by tDMRG truncation errors. 

\begin{figure}[H]
  \centering
  \hspace*{-0.2cm}
  \includegraphics[width=0.48\textwidth]{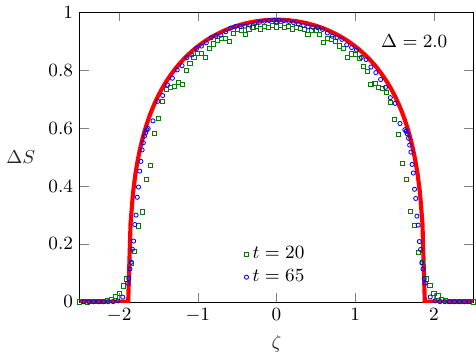}
  \hspace*{0.2cm}
  \includegraphics[width=0.48\textwidth]{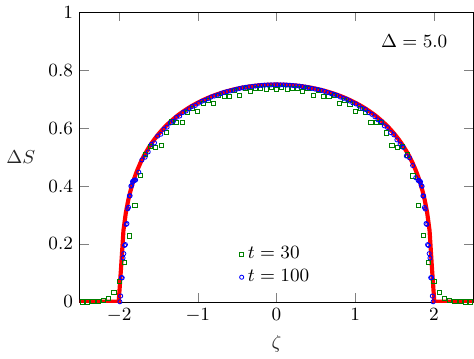}\\
  \hspace*{-0.2cm}
  \includegraphics[width=0.48\textwidth]{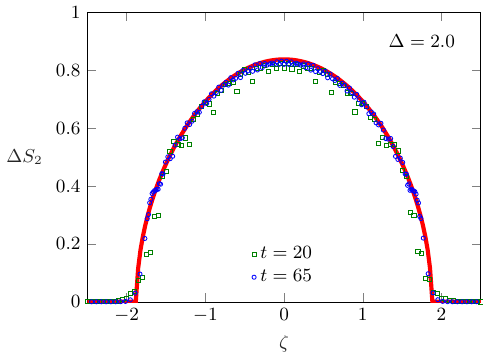}
  \hspace*{0.2cm}
  \includegraphics[width=0.48\textwidth]{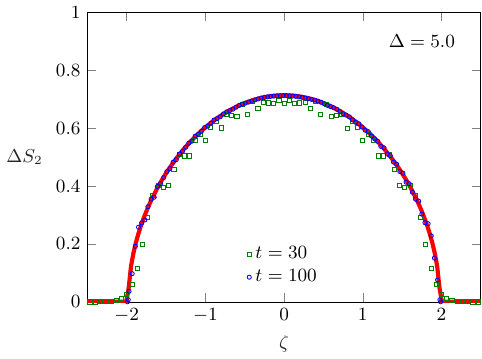}  
  \caption{Entropy profiles $\Delta S$ (top) and $\Delta S_2$ (bottom) after a domain-wall
  excitation $\mtop_1$ for two different values of $\Delta$ and $\chlength = 400$.
  The solid lines show the ansatz Eq.~\eqref{eq: Sb1} for the von Neumann, as well as
  Eq.~\eqref{eq: Sbn} for the $n=2$ R\'enyi excess entropy.}
  \label{fig: S1S2}
\end{figure}

Although we find a nontrivial asymptotic behaviour of the half-chain entanglement,
we expect that the full profile should still be described, up to a multiplicative factor,
by the quasiparticle ansatz introduced in section \ref{sec: QP}, similarly to the
Majorana excitation in the gapless phase in Fig. \ref{fig: JW_S1_allbond}.
Therefore, we put forward the ansatz
\begin{equation}
    \Delta S = \left(1+\frac{S(0)}{\ln 2}\right) \left[ - \mathcal{N} \ln \left( \mathcal{N} \right) - \left(1-\mathcal{N} \right) 
    \ln \left( 1-\mathcal{N} \right) \right]\, ,
    \label{eq: Sb1}
\end{equation}
and for the excess R\'enyi entropy we propose
\begin{equation}
    \Delta S_n = \left(1+\frac{S_n(0)}{\ln 2}\right) \frac{1}{1-n} \ln \left[ \, \mathcal{N}^n + (1-\mathcal{N})^n \right]\, .
    \label{eq: Sbn}
\end{equation}
The quasiparticle fraction $\mathcal{N}$ must now be evaluated via \eqref{eq: N}
by using the spinon velocities \eqref{eq: vs_gapped} in the gapped phase.
Note that the binary entropy functions are multiplied by a factor to reproduce
our findings for the half-chain, where $\mathcal{N}=1/2$.
The results of our numerical calculations for the profiles $\Delta S$ and $\Delta S_2$,
plotted against the scaling variable $\zeta=r/t$, are shown in Fig.~\ref{fig: S1S2}.
The solid lines show the respective ansatz \eqref{eq: Sb1} and \eqref{eq: Sbn},
which give a very good description of the data for both $\Delta$ values shown.
In fact, we checked that the profiles are nicely reproduced even for $\Delta=1.5$,
which already corresponds to a relatively large correlation length. 

%
%

\subsection{Magnetization profiles}

To conclude this section, we also investigate the spreading of the magnetization profiles
for the antiferromagnetic domain wall excited by $\mtop_1$. This setting was studied
previously with a focus on the edge behaviour of the profile \cite{zauner2015comoving}.
In order to remove the dependence on the ground-state value \eqref{eq: sz_staggered}
of the staggered magnetization, we consider the normalized profile
\begin{equation}
    \mathcal{M}_j(t) = 
    \frac{ \bra{\psi(t)} \sigma^z_j \ket{\psi(t)} }{ \bra{\psi_\uparrow} \sigma^z_j \ket{\psi_\uparrow} },
    \label{eq: M_j}
\end{equation}
which then varies between $-1\le  \mathcal{M}_j(t) \le 1$ along the chain.
We are mainly interested in the quasiparticle description of the time-evolved profile.
In fact, a very similar problem was studied for a ferromagnetic domain wall in the XY chain \cite{eisler2020frontdyn},
by first expanding the excited state in the single-particle basis of the Hamiltonian,
which can then be time evolved trivially. 

Here we assume that the dominant weight for our simple domain wall is carried by
single-spinon excitations $\ket{q}$.
Strictly speaking, this is only possible if one considers antiperiodic or open boundary
conditions on the spins, since for a periodic chain spinons are created in pairs
(i.e. one actually has a pair of domain walls). The time evolved state can then be written as
\begin{equation}
    \ket{\psi(t)} \simeq
    \sum_q \ee^{-it\varepsilon_s(q)} c(q) \ket{q}\, ,
    \label{eq: psit1s}
\end{equation}
where $\varepsilon_s(q)$ is the spinon dispersion \eqref{eq: E_s2},
while $c(q)$ are the overlaps of the domain-wall excitation with the single-spinon states.
Note that the momentum of a single spinon satisfies $0\le q \le\pi$,
however, the total momentum of spinons above the quasidegenerate ground state is shifted by $\pi$.
Since the domain wall is created by a strictly local fermionic operator, we assume that
in the thermodynamic limit $|c(q)|$ becomes a constant in momentum space,
i.e. $c(q)=\ee^{i\alpha(q)}/\sqrt{N}$ is just a phase factor normalized by the number $N$ of spinon states.
Using this in \eqref{eq: psit1s}, one obtains for the profile
\begin{equation}
    \mathcal{M}_j(t) = 
    \frac{1}{N}\sum_p \sum_q \ee^{-it(\varepsilon_s(q) - \varepsilon_s(p))} \ee^{i(\alpha(q)-\alpha(p))}
     \frac{\bra{p} \sigma^z_j\ket{q}}{\bra{\psi_\uparrow} \sigma^z_j \ket{\psi_\uparrow} }\, .
    \label{eq: M_j2}
\end{equation}

Clearly, the main difficulty of calculating \eqref{eq: M_j2} is due to the form factors $\bra{p} \sigma^z_j\ket{q}$.
For the transverse Ising and XY chains, such form factors are known explicitly \cite{iorgov2011IsingFF,iorgov2011XYFF} and were used
to obtain a double integral representation of the magnetization profile \cite{eisler2016universal,eisler2020frontdyn}.
The hydrodynamic limit can then be obtained from the stationary-phase analysis of the integrals.
Moreover, there exists a number of form factor results for the XXZ chain as well (see e.g.
\cite{kitanine1999formfactor,dugave2015formfactor}), which were used in numerical studies
of the magnetization profile after a spin-flip excitation \cite{vlijm2016spinondyn}. 
Unfortunately, however, the expressions are typically rather involved or not in a representation
that could be useful for our purposes. In fact, we are not aware of any results where the required
single-spinon matrix elements are evaluated as a function of the spinon rapidity or momentum.

Nevertheless, based on the known results, we give a handwaving argument about how
the main structure of the form factor should look like. Most importantly, we assume that it becomes singular
for $p \to q$ and can be written as
\eq{
\lim_{p \to q} \frac{\bra{p} \sigma^z_j\ket{q}}{\bra{\psi_\uparrow} \sigma^z_j \ket{\psi_\uparrow}}
\simeq \frac{i}{N} \ee^{i(q-p)j}\frac{\mathcal{F}(q)}{p-q} \, .
}
Here the only $j$-dependence is in the exponential factor that follows from the action of the translation operator,
and the function $\mathcal{F}(q)$ denotes the slowly varying part of the form factor around its pole.
The factor $1/N$ is required for a proper thermodynamic limit of \eqref{eq: M_j2}.
Converting the sums into integrals, one can proceed with the stationary phase analysis
similarly to the XY case \cite{eisler2020frontdyn}, by expanding the phases around $Q=q-p=0$.
Using a resolution of the pole and the definition of the step function
\eq{
\frac{1}{Q}=i \pi \delta(Q) + \lim_{\epsilon\to 0}\frac{1}{Q+i\epsilon}, \qquad
\Theta(x) = -\lim_{\varepsilon \to 0} \int_{-\infty}^{\infty} \frac{\dd Q}{2\pi i}
\frac{\ee^{-iQx}}{Q+i\varepsilon},
}
one arrives at the following simple expression for the profile
\begin{equation}
    \mathcal{M}_j(t) = 1 - 2 \, \tilde{\mathcal{N}}\, , \qquad
    \tilde{\mathcal{N}}  = \int_{0}^{\pi} \frac{\dd q}{\pi} \,
    \Theta( v_s(q)t - j ) \, \mathcal{F}(q)\, .
    \label{eq: M_j3}
\end{equation}
Note that the proper normalization of the profile for $t=0$ requires to have
\eq{
\int_{0}^{\pi} \frac{\dd q}{\pi} \, \mathcal{F}(q)=1\, .
    \label{eq: intF}
}

\begin{figure}[H]
  \centering
  \hspace*{-0.2cm}
  \includegraphics[width=0.48\textwidth]{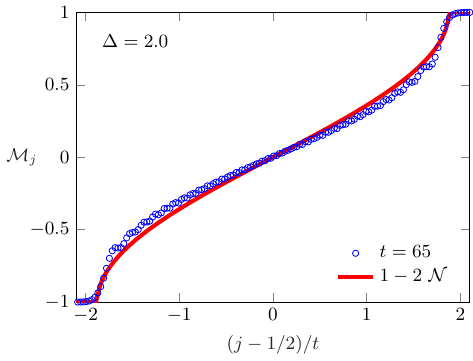}
  \hspace*{0.2cm}
  \includegraphics[width=0.48\textwidth]{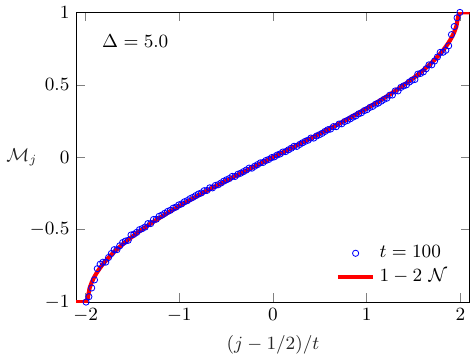}
   \caption{Normalized magnetization profiles $\mathcal{M}_j(t)$ obtained from tDMRG calculations for
   $\Delta = 2.0$ (left) and $\Delta = 5.0$ (right) after a domain-wall excitation $\mtop_1$ in a chain of
   length $\chlength = 400$. The solid lines show the ansatz $1 - 2 \, \mathcal{N}$, with the spinon fraction
   Eq.~\eqref{eq: N} calculated from the velocities in Eq.~\eqref{eq: vs_gapped}.}
   \label{fig: mag}
\end{figure}

The result \eqref{eq: M_j3} is nothing else but the quasiparticle interpretation
of the magnetization profile in the hydrodynamic limit. Indeed, the initial sharp domain wall
is carried away by spinons of different momenta $q$ and velocities $v_s(q)$,
where $\mathcal{F}(q)$ gives the corresponding weight. Unfortunately, without
an explicit analytical result on the form factor, one has to make a guess on the
weight function. The simplest assumption is $\mathcal{F}(q)\equiv1$, which indeed holds
true for the XY chain form factors \cite{eisler2020frontdyn}. With this simple choice
one actually has $\tilde{\mathcal{N}}=\mathcal{N}$, that is we recover the spinon
fraction introduced in \eqref{eq: N} for the description of the entropy profile.
In Fig.~\ref{fig: mag} we show the comparison of this simple ansatz to the
tDMRG data, with a rather good agreement for a large $\Delta=5$.
For $\Delta=2$, however, one can already see the deviations from our
simple ansatz, which fails completely for even smaller anisotropies.
Thus, in sharp contrast to the case of the entanglement entropies,
the spinon contributions to the magnetization cannot be taken to be equal,
except for close to the Ising limit.

%
%

\section{Summary and discussion \label{sec: summary}}

We studied the entanglement spreading in the XXZ chain after excitations that are strictly
local in terms of the fermion operators. In the gapless phase we found that the time evolution after
a fermion creation operator yields an entropy profile that can be well described by a probabilistic
quasiparticle ansatz for not too large $\Delta$, assuming equal contributions from low-lying spinon excitations.
On the other hand, for a local Majorana excitation we observe that the quasiparticle ansatz
holds only up to a multiplicative factor, determined by the excess entropy at the operator insertion point.
This is in agreement with our CFT calculations, which suggest that the excess entropy exceeds
$\ln(2)$ for any $\Delta\ne0$, with a very slow convergence towards the asymptotic
value $2\ln(2)$. In the symmetry-broken gapped phase we considered a different
Majorana excitation, creating an antiferromagnetic domain wall. For the entropy profile
we find again a nontrivial prefactor, whereas our simple ansatz for the magnetization,
assuming equal contributions from the spinons, holds only in the Ising limit $\Delta\to \infty$.

The main limitation of our quasiparticle ansatz \eqref{eq: delS_N} is that it includes only
the low-lying spinons. It is natural to ask how well such an assumption actually holds for
our local excitations in the different regimes. A simple way to quantify the spectral weight
of the spinons in the gapless regime is via the energy difference
$\braket{\Delta E}=\bra{\psi_0} (m_1 H m_1-H)\ket{\psi_0}$
of the Majorana excitation (equal to that of $c_1^\dag$ by particle-hole symmetry)
measured from the ground state,
whereas in the gapped case we replace $m_1 \to \tilde m_1$.
Our assumption in both regimes was that one can practically work with single-spinon states,
whose energies above the ground state are given by the corresponding dispersions $\varepsilon_s(q)$
in \eqref{eq: E_s} and \eqref{eq: E_s2}, respectively. This yields the simple formula
for the energy difference
\begin{equation}
    \expval{\Delta E} = \int_{0}^\pi \varepsilon_s(q) \frac{\dd q}{\pi}\, .
    \label{eq: delE}
\end{equation}

\begin{figure}[H]
  \centering
    \includegraphics[width=0.48\textwidth]{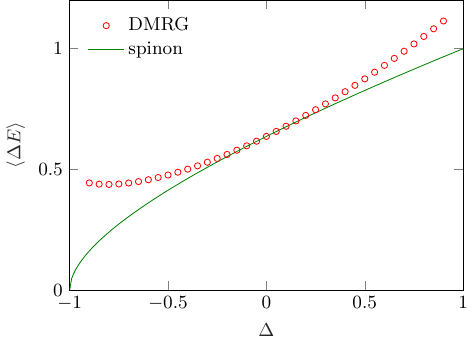}
    \includegraphics[width=0.48\textwidth]{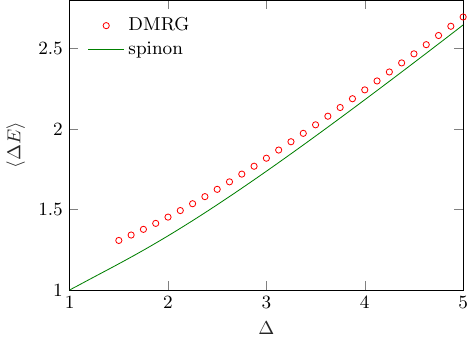}
   \caption{Energy difference due to the insertion of local operator $m_1$ in the gapless (left)
   and $\tilde m_1$ in the gapped (right) regime.
   DMRG results (symbols) for $\chlength = 400$ are compared to the spinon ansatz (lines) in Eq.~\eqref{eq: delE}.
   Note the different vertical scales.}
   \label{fig: Egap}
\end{figure}

To test the validity of our assumption, in Fig.~\ref{fig: Egap} we compare the energy difference
obtained from DMRG to the formula \eqref{eq: delE} in both gapless and gapped phases.
As expected, the result at the free-fermion point $\Delta=0$ is exactly reproduced, while
the error remains relatively small in the regime $|\Delta| \lesssim 0.5$. However, not surprisingly,
the overall behaviour of $\braket{\Delta E}$ is not properly captured by the naive ansatz
\eqref{eq: delE}, especially for $\Delta \to -1$, which is exactly what we observed for the entropy
profiles in Fig. \ref{fig: tDMRG_vs_QPk}. On the other hand, in the gapped phase shown on the
right of Fig.~\ref{fig: Egap}, one has a qualitatively good description in the entire regime,
with the error decreasing for $\Delta \gg 1$. This explains why we had a much better
overall description of the entropy profiles for $\Delta>1$ via the quasiparticle ansatz \eqref{eq: Sb1}.
 
%
%

Another feature that is not completely understood is the multiplicative factor
of the spinon ansatz appearing for Majorana excitations. In the gapless phase this
could be accounted for the mixing of the chiral boson modes and yields a factor $2$
in the limit $t\to\infty$ for any $\Delta \ne 0$. The exceptional behaviour of the
XX chain can actually be also understood directly, using a duality transformation 
\cite{perk1977duality,perk1977duality2,peschel1984duality,turban1984duality}
that relates it to two independent and critical transverse Ising chains.
Furthermore, as shown in \cite{eisler2016universal}, the Majorana excitation on the
XX chain transforms under the dual map into a Majorana excitation acting only on
a \emph{single} Ising chain. Hence, the asymptotic excess entropy is given by $\ln(2)$
and there is no doubling in this case.
On the other hand, in the gapped phase the prefactor in \eqref{eq: Sb1} seems to be
nontrivially related to the ground-state entanglement entropy. Note that a similar observation
was reported after a local quench in the non-critical transverse Ising chain \cite{eisler2008entev},
where the entanglement plateau was also found to be related to the ground-state value.
A deeper understanding of these effects requires further studies.

Finally, let us comment about the case where the locality of the excitation
is not imposed in the fermionic but rather in the spin picture. In other words,
instead of the $c^\dag_j$ excitation one could consider the spin operator
$\sigma^+_j$ by dropping the Jordan-Wigner string in \eqref{eq: fop}.
According to our tDMRG calculations carried out for this case, the entropy
profiles change completely, becoming more flat in the center with a maximum that
stays way below $\ln(2)$. In short, the fermionic nature of the local excitations
turns out to be essential for the applicability of the quasiparticle description.

\section*{Acknowledgements}

The authors acknowledge funding from the Austrian Science Fund (FWF) through
project No. P30616-N36.

\begin{appendix}

\section{Correlation functions of vertex operators\label{sec: CFT_appendix}}

In the following we give the main steps of the calculation of the excess R\'enyi entropy
$\Delta S_2$, obtained via the ratio \eqref{eq: DSn} of four-point and two-point functions.
 As in the main text, we consider two different local operators, the one corresponding
 to the fermion creation
\eq{
\mathcal{O}_f=
\ee^{ik_Fd} \psi^\dag + \ee^{-ik_Fd} \bar\psi^\dag \, ,
}
as well as the Hermitian Majorana excitation
\eq{
\mathcal{O}_m=
\ee^{ik_Fd} \psi^\dag + \ee^{-ik_Fd} \bar\psi^\dag +
\ee^{-ik_Fd} \psi + \ee^{ik_Fd} \bar\psi \, .
\label{eq: Omapp}}
They are composed of chiral fermion fields which, after the Bogoliubov
transformation \eqref{eq: bt}, can be written as vertex operators \eqref{eq: vop}
involving chiral boson fields. The holomorphic and anti-holomorphic components
of the vertex operators are summarized in the table below, where $c=\cosh(\xi)$
and $s=\sinh(\xi)$.
\begin{table}[h]
\begin{center}
\begin{tabular}{ | l | c | c | c | r |}
\hline
   & $\psi$ & $\psi^\dag$ & $\bar\psi^{\phantom{\dag}}$ & $\bar\psi^\dag$  \\
\hline
  $\alpha$ & $-c$ & $c$ & $s$ & $-s$ \\
\hline
  $\beta$  & $-s$ & $s$  & $c$ & $-c$ \\
\hline
\end{tabular}
\end{center}
\caption{Parameters of the vertex operators \eqref{eq: vop} for the fermionic fields}
\end{table}

We start by evaluating the two point function in the denominator of \eqref{eq: DSn}.
Using the fact that vertex operators are primaries with conformal dimensions
$h=\alpha^2/2$ and $\bar h=\beta^2/2$, one immediately obtains the nonvanishing
two-point functions on the plane as
\eq{
\begin{aligned}
\braket{\psi(w_1,\bar w_1)\psi^\dag(w_2, \bar w_2)} \propto
(w_1-w_2)^{-c^2} \, (\bar w_1- \bar w_2)^{-s^2}, \\
\braket{\bar \psi(w_1, \bar w_1)\bar \psi^\dag(w_2, \bar w_2)} \propto
(w_1-w_2)^{-s^2} \, (\bar w_1- \bar w_2)^{-c^2}.
\end{aligned}
}
From \eqref{eq: w12} we have $w_1- w_2=\bar w_1 - \bar w_2=2\epsilon$, thus
we obtain for the two-point functions
\eq{
\braket{\mathcal{O}^\dag_f (w_1,\bar w_1) \mathcal{O}_f (w_2,\bar w_2)} =
2\, (2\epsilon)^{-(c^2+s^2)}, \qquad
\braket{\mathcal{O}^\dag_m (w_1,\bar w_1) \mathcal{O}_m (w_2,\bar w_2)} =
4\, (2\epsilon)^{-(c^2+s^2)}.
}

Let us now move to the four-point function on the Riemann surface $\Sigma_2$.
This is a sum of many terms, from which the nonvanishing contributions allowed by the
neutrality conditions \eqref{eq: neut} are given by
\eq{
\braket{\psi \psi^\dag \bar \psi \bar \psi^\dag}, \quad
\braket{\bar \psi \bar \psi^\dag \psi \psi^\dag }, \quad
\braket{\bar \psi \psi^\dag \psi \bar \psi^\dag }, \quad
\braket{\psi \bar \psi^\dag \bar \psi \psi^\dag }, \quad
\braket{\psi \psi^\dag \psi \psi^\dag}, \quad
\braket{\bar \psi \bar \psi^\dag \bar \psi \bar \psi^\dag}.
\label{4pterms}}
We first analyze the Jacobian of the transformation \eqref{eq: conf_trans} from $\Sigma_2\to\Sigma_1$.
The derivatives of the mapping are given by
\begin{equation}
    \frac{\dd w}{\dd z} = i \ell \frac{n z^{n-1}}{(1-z^n)^2}\, , \qquad
    \frac{\dd \bar{w}}{\dd \bar{z}} = -i \ell \frac{n \bar{z}^{n-1}}{(1-\bar{z}^n)^2}\, .
\end{equation}
Introducing the variable
\eq{
\chi = \frac{(1-z_1^2)^2(1-z_2^2)^2}{4z_1 z_2},
\label{chi}}
one obtains for the first four contributions in \eqref{4pterms}
\eq{
\ell^{-2(c^2+s^2)}\chi^{c^2/2} \bar \chi^{s^2/2} \chi^{s^2/2} \bar \chi^{c^2/2}=
\ell^{-2(c^2+s^2)}|\chi|^{c^2+s^2},
}
whereas for the last two contributions we have, respectively
\eq{
\ell^{-2(c^2+s^2)}\chi^{c^2} \bar \chi^{s^2}, \qquad
\ell^{-2(c^2+s^2)}\chi^{s^2} \bar \chi^{c^2}.
}
Note that there is an extra sign factor
$(-i)^{c^2}(i)^{s^2}(i)^{s^2}(-i)^{c^2}=(-i)^{2(c^2-s^2)}=-1$ which multiplies
the first two Jacobian.

The next step is to evaluate the vertex four-point functions.
Using \eqref{eq: vertexcorr} this reads for the first term in \eqref{4pterms}
\eq{
z_{12}^{-c^2} z_{34}^{-s^2}
z_{13}^{-cs} z_{24}^{-cs}
z_{14}^{cs} z_{23}^{cs}
\bar z_{12}^{-s^2} \bar z_{34}^{-c^2}
\bar z_{13}^{-cs} \bar z_{24}^{-cs}
\bar z_{14}^{cs} \bar z_{23}^{cs}=
(-1)
|1-\eta|^{2cs}|\eta|^{-(c^2+s^2)}
|z_{13}z_{24}|^{-(c^2+s^2)}
\label{vc1}}
Note that we have used the property $z_{34}=-z_{12}$.
It is easy to check that one obtains the very same factor from the second term.
Similarly, using $z_{23}=z_{14}$, one can check that the third and fourth terms deliver
\eq{
z_{12}^{cs} z_{34}^{cs}
z_{13}^{-cs} z_{24}^{-cs}
z_{14}^{-c^2} z_{23}^{-s^2}
\bar z_{12}^{cs} \bar z_{34}^{cs}
\bar z_{13}^{-cs} \bar z_{24}^{-cs}
\bar z_{14}^{-s^2} \bar z_{23}^{-c^2}=
|\eta|^{2cs}|1-\eta|^{-(c^2+s^2)}
|z_{13}z_{24}|^{-(c^2+s^2)}.
\label{vc2}}
For the fifth term one has
\eq{
\left[\eta(1-\eta)\right]^{-c^2} (z_{13}z_{24})^{-c^2}
\left[\bar \eta(1- \bar \eta)\right]^{-s^2} (\bar z_{13} \bar z_{24})^{-s^2} \, ,
\label{vc3}}
and the last term follows by interchanging $c$ and $s$ above.

In order to obtain an expression in terms of the cross-ratios, one can rewrite \eqref{chi} as
\eq{
\chi = \left(\frac{\ell}{2\epsilon}\right)^2 \, \eta(1-\eta) \, z_{13}z_{24} \, .
\label{chi2}}
Putting everything together, one arrives at the four-point function
\eq{
2 \, (2\epsilon)^{-2(c^2+s^2)}
\left[|\eta|^{(c+s)^2} + |1-\eta|^{(c+s)^2} + 1 \right].
}

Evaluating the four-point function for the Majorana excitation \eqref{eq: Omapp}
is more cumbersome, since one has
a large number of terms to consider. There are, however, some simple
rules and symmetry arguments which make the task easier.
First of all, one should clearly always have the same number of
creation and annihilation operators, for the neutrality conditions \eqref{eq: neut}
of the vertex correlation functions to be satisfied. This already
drastically reduces the number of terms to consider.
The remaining ones can be collected into families, some of them
given by \eqref{4pterms}.

Let us consider the family generated by the first term in \eqref{4pterms},
by allowing permutations of the left- and right-moving operators separately
(i.e. interchanging the first or last two operators).
If only the first or last two are interchanged, the vertex correlator \eqref{vc1}
is modified by replacing
\eq{
|1-\eta|^{2cs} \to |1-\eta|^{-2cs} \, ,
}
whereas the correlator remains the same if both of them are interchanged.
The next family is generated by the second term in \eqref{4pterms}, which
is actually related to the first one by Hermitian conjugation. Hence this
just gives a factor of two. The same argument holds for the next two families,
where interchanging only one pair modifies the correlator in \eqref{vc2} as
\eq{
|\eta|^{2cs} \to |\eta|^{-2cs} \, .
}
Finally, the single interchange in the fifth family leads to
\eq{
(1-\eta)^{-c^2} \to (1-\eta)^{c^2} , \qquad
(1-\bar\eta)^{-s^2} \to (1-\bar\eta)^{s^2} , \qquad
}
whereas the last family follows by interchanging $c$ and $s$ above.

There are, however, two additional families appearing where the left-
and right-moving particles are intertwined. They are given by the
representative correlators
\eq{
\braket{\psi \bar\psi^\dag \psi^\dag \bar\psi}, \quad
\braket{\bar\psi \psi^\dag \bar\psi^\dag \psi}.
}
Defining the variable
\eq{
\sigma = \frac{(1-z_1^2)^2(1-\bar z_2^2)^2}{4z_1 \bar z_2},
}
the corresponding Jacobians contain the factors
$\sigma^{c^2} \bar\sigma^{s^2}$ and $\sigma^{s^2} \bar\sigma^{c^2}$, respectively.
Furthermore, the vertex correlation functions yield
\eq{
|\eta|^{\pm2cs}|1-\eta|^{\mp2cs}
(z_{13} \bar z_{24})^{-c^2}(\bar z_{13}z_{24})^{-s^2}, \qquad
|\eta|^{\pm2cs}|1-\eta|^{\mp2cs}
(z_{13} \bar z_{24})^{-s^2}(\bar z_{13}z_{24})^{-c^2},
}
and each term comes with a double multiplicity. Collecting all the terms, the four-point
function takes the form
\eq{
2 \, (2\epsilon)^{-2(c^2+s^2)} (2A + B + C) \, ,
}
where the factors $A$, $B$ and $C$ are reported in \eqref{eq: A}-\eqref{eq: C}.

\end{appendix}

\bibliography{literatur.bib}

\nolinenumbers

\end{document}